%% file: main.tex
\documentclass[%
reprint,
superscriptaddress,
amsmath,amssymb,
aps,
pre,
]{revtex4-2}

\usepackage[german,english]{babel}
\usepackage{graphicx}
\usepackage{mathtools} 
\usepackage{siunitx} 
\usepackage{bbold} 

\newcommand*\diff{\mathop{}\!\mathrm{d}}

\newcommand{\Rho}{\mathrm{P}} 

\usepackage{dcolumn}
\usepackage{bm}

\usepackage{tabularx} 
\newcolumntype{G}{X}
\newcolumntype{P}{>{\hsize=.5\hsize}X}

\usepackage{afterpage}

\begin{document}
	
	
	\title{Energetic constraints on filament mediated cell polarization}
	
	\author{Harmen Wierenga}
	\affiliation{AMOLF, Science Park 104, 1098 XG Amsterdam, The Netherlands}
	\author{Pieter Rein ten Wolde}
	\email{p.t.wolde@amolf.nl}
	\affiliation{AMOLF, Science Park 104, 1098 XG Amsterdam, The Netherlands}

	
	\begin{abstract}
		    Cell polarization underlies many cellular processes, such as differentiation, migration, and budding.
		    Many living cells, such as budding yeast and fission yeast,
	        use cytoskeletal structures to actively transport proteins to one location on the membrane
	        and create a high density spot of membrane-bound proteins.
	        Yet, the thermodynamic constraints on filament-based cell polarization remain unknown.
	        We show by mathematical modeling that cell polarization requires detailed balance to be broken,
	        and we quantify the free-energy cost of  maintaining a polarized state of the cell.
	        Our study reveals that detailed balance cannot only be broken via the active transport of proteins along filaments,
	        but also via a chemical modification cycle, allowing detailed balance to be broken
	        by the shuttling of proteins between the filament, membrane, and cytosol.
	        Our model thus shows that cell polarization can be established via two distinct driving mechanisms,
	        one based on active transport and one based on non-equilibrium binding.
	        Furthermore, the model predicts that the driven binding process dissipates orders of magnitude less free-energy
	        than the transport-based process to create the same membrane spot.
	        Active transport along filaments may be sufficient to create a polarized distribution of membrane-bound proteins,
	        but an additional chemical modification cycle of the proteins themselves is more efficient
	        and less sensitive to the physical exclusion of proteins on the transporting filaments,
	        providing insight in the design principles of the Pom1/Tea1/Tea4 system in fission yeast and the Cdc42 system in budding yeast.
	\end{abstract}
	
	\pacs{Valid PACS appear here}
	\maketitle
	

Cell polarization is a common motif for establishing different cellular functions and for cell development,
in which a cell generates a distinct front and back.
For example, cells that perform unidirectional movement need to polarize along a single axis~\cite{campanale_development_2017}.
Lophotrichous bacteria require the placement of multiple flagella on one side of the cell,
and crawling eukaryotic cells need to polarize their cytoskeleton to create a protrusive leading edge on one side
and a contractile trailing edge on the opposite side of the cell~\cite{li_beyond_2008,dogterom_actinmicrotubule_2019}.
Moreover, epithelial cells are polarized to distinguish the apical and basal sides~\cite{li_beyond_2008,campanale_development_2017},
and asymmetric cell division requires cell polarization along the division axis
to create different fates for the daughter cells~\cite{pillitteri_asymmetric_2016,campanale_development_2017}.
For instance, budding yeast requires the formation of a bud on one spot on its cell membrane~\cite{martin_cell_2014}.
Similarly, fission yeast remains polar after cell division, primarily growing at the old pole initially~\cite{martin_cell_2014}.

Because cell polarization is an essential cellular feature,
many different biological processes exist that induce cell polarization~\cite{rappel_mechanisms_2017}.
A large class of such processes involves the cytoskeletal filaments~\cite{raman_polarized_2018},
because both microtubules and actin filaments have an intrinsically polar structure
by which they can act as tracks for the directional transport of cargoes by motor proteins.
Because the cytoskeleton itself is often asymmetrically organised, for example in the mitotic spindle,
these structures can be used to guide other proteins into a polarized state~\cite{li_beyond_2008}.
For example, motor proteins that walk on central spindle microtubules can transport proteins
such as the RHO activator ECT2~\cite{dogterom_actinmicrotubule_2019} towards the membrane,
where they can promote the formation of the cytokinetic ring.
In budding yeast, the small GTPase of the Rho family Cdc42 is bound to the membranes of vesicles
that are delivered to the membrane along actin cables~\cite{harris_cdc42_2010,martin_cell_2014},
which may produce cell polarization.
Furthermore, fission yeast uses microtubule based transport to place the proteins Tea4 and Tea1 at the membrane,
where Tea4 forms a complex with Dis2 and dephosphorylates the DYRK family kinase Pom1,
which subsequently binds to the membrane~\cite{hachet_phosphorylation_2011,martin_cell_2014}.
Once Pom1 is on the membrane, it autophosphorylates and unbinds again,
leaving a steady state distribution of Pom1 on the membrane in the neighbourhood of the microtubule tips~\cite{hachet_phosphorylation_2011}.

The transport of vesicle-bound Cdc42 in budding yeast shows that active transport can play a role in
creating a high density of proteins in one spot on the membrane~\cite{harris_cdc42_2010,martin_cell_2014}.
In contrast, the organism fission yeast does not directly transport Pom1, but it uses (de)phosphorylation to drive the protein through a chemical modification cycle
where it binds to the membrane preferably near the positions of the microtubule tips
where Dis2 is present~\cite{hachet_phosphorylation_2011,martin_cell_2014}.
Here, we investigate which of these two mechanisms, active transport along a filament or the chemical driving of a binding cycle catalysed by the cytoskeleton,
is more efficient in creating a polarized distribution of proteins on the membrane.
Because a polarized state corresponds to a non-equilibrium distribution of the protein,
the maintenance of this distribution requires the constant dissipation of chemical free energy, usually in the form of NTP hydrolysis.
To assess the efficiency of both active transport and driven chemical modification in creating a polarized protein distribution,
we take into account the chemical free-energy dissipation of each process.
Such energetic constraints are typically excluded when discussing cellular pattern formation,
but they are important because they provide a quantitative measure by which we can compare different mechanisms for polarising a cell.
Using a minimal model in which both transport along a filament and non-equilibrium binding can lead to cell polarization,
we will show that transport alone can be sufficient for creating a polarized spot on the membrane,
but that a chemical modification cycle of the protein itself can dissipate
orders of magnitude less free energy to achieve the same quality of polarization.
This may explain why many cell polarization systems, including the Cdc42 system in budding yeast and the Pom1 system in fission yeast,
contain a chemical modification cycle.

\section{\label{sec:polarisation_model}Minimal model for membrane spot formation}
\begin{figure}
	\includegraphics{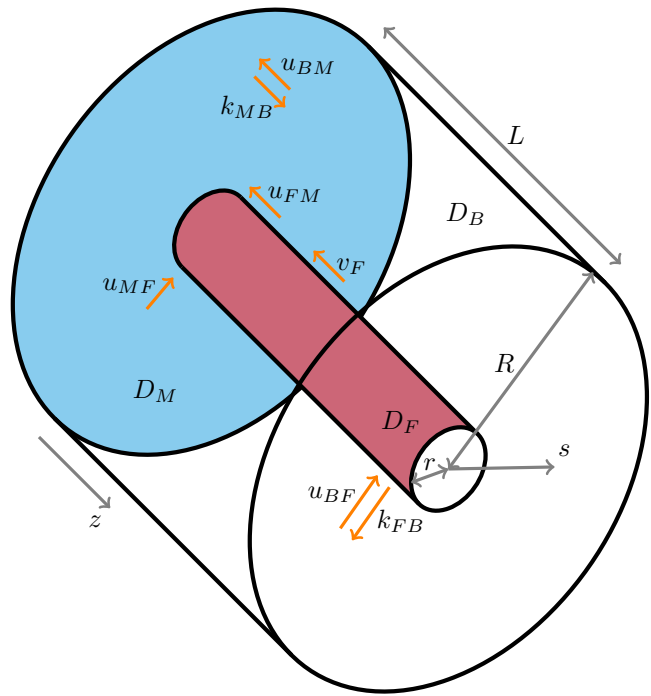}
	\caption{\label{fig:polarisation_model}
		Minimal model to investigate how protein transport along a filament can create
		a polarized distribution of membrane bound proteins.
		We consider a subvolume of a cell in the shape of a closed cylinder with radius $R$,
		with a circular patch of the membrane on one side (blue).
		A finite filament (red) with a radius $r$ and a length $L$ is located in the center of the cylinder, perpendicular to the membrane.
		We consider a single protein species that can be membrane bound ($M$),
		dissolved in the bulk ($B$), or connected to the filament by motor proteins ($F$).
		In each of these parts, the particles diffuse with diffusion constants $D_{M}$, $D_{B}$, and $D_{F}$, respectively,
		and on the filament the motor proteins provide an average drift velocity $v_{F}$ towards the membrane.
		The particles can unbind from the membrane and the filament with the Markovian rates $k_{MB}$ and $k_{FB}$.
		The corresponding reverse binding transitions can occur when a particle that is dissolved in the bulk
		is in contact with the membrane or the filament, and the transition rates $u_{BM}$ and $u_{BF}$ have the dimensions of a velocity, not of a rate.
		Finally, a particle that is bound to the filament and in contact with the membrane can transition to the membrane with a rate $u_{FM}$
		that is possibly different from $u_{BM}$,
		and similarly the particle can bind from the membrane to the filament with a rate $u_{MF}$ that can be different from $u_{BF}$.
		We define a coordinate system with $z$ longitudinal to the filament, such that the membrane is at $z=0$,
		and with $s$ as the radial distance, with the microtubule surface at $s=r$.
	}
\end{figure}

Inspired by the cell polarization systems in fission yeast and budding yeast,
we create an analytically solvable minimal model in which a filament can transports proteins
and where the proteins bind to a membrane patch at the end of the filament.
We consider a single stable filament that is perpendicular to the cell membrane, as shown in Fig.\,\ref{fig:polarisation_model}.
All model parameters are shown and explained in Fig.\,\ref{fig:polarisation_model}.
We do not model the entire cell, but only a finite volume around the filament,
because diffusion will smooth out the protein distributions sufficiently far from the filament.
By choosing a cylindrical shape for this volume, we can find analytical solutions for the probability densities in each part of the system.

The membrane patch is located on one side of the cylinder, so the proteins can be on the membrane, on the filament, or in the cytosol (bulk).
Crucially, the particles are transported along the filament towards the membrane,
and because the filament is in direct contact with the membrane, particles can transition between the membrane and the filament.
Additionally, we include transitions between bulk and the filament and between the bulk and the membrane.
Hence, a particle flux can exist that moves from the bulk along the filament to the membrane, and from there back to the bulk.
The transition pathways between the three system parts are memoryless,
and we include the microscopic reverse reaction for each transition such that the free-energy dissipation in the system remains finite.
These transitions can be out of equilibrium by coupling the chemical reactions to a non-equilibrium NTP bath,
such that the binding rates alone break detailed balance and cause a flux of proteins
from the bulk to the filament, to the membrane, and back to the bulk again.
This mechanism can act independently from and concurrently with active transport to create a polarized distribution of membrane-bound proteins.
In the following sections, we will first keep the binding reactions in equilibrium to assess how efficiently the non-equilibrium distribution
on the membrane is maintained by active transport along the filament.
Subsequently, we will also take into account the effects of a non-equilibrium binding cycle.

We use cylindrical coordinates $(s,z)$, where the radial distance $s$ runs from the filament radius $r$ to the container radius $R$,
and the longitudinal coordinate $z$ runs from $0$ to the filament length $L$.
Because the system is rotationally symmetric, the resulting concentration profiles will be too,
and we do not require the azimuth.
Having a finite microtubule radius eases the mathematics of membrane-filament transitions compared to a one dimensional microtubule,
by avoiding unphysical divergences of the protein densities.
We model the dynamics of the system in time $t$ by a set of coupled Fokker-Plank equations for the protein densities
$f(z,t)$, $m(s,t)$ and $b(z,s,t)$, which cover the filament, the membrane and the bulk, respectively.

To make the equations analytically tractable, we first consider a system with a large bulk diffusion constant $D_{B}\rightarrow \infty$,
which smears out density fluctuations and turns the bulk field into a single concentration,
\begin{equation}
	b\!\left(z,s,t\right) = b\!\left( t \right).
	\label{eq:bulk_scalar}
\end{equation}
The partial differential equations for the protein concentrations read
\begin{align}
	\partial_{t} f \left( z,t \right) &= D_{F} \partial_{z}^{2} f \left( z,t \right) + v_{F} \partial_{z} f \left( z,t \right) \nonumber \\*
	&\quad- k_{FB} f \left( z,t \right) + 2 \pi r u_{BF} b \left( t \right), \label{eq:time_evol_f} \\
	\partial_{t} m \left( s,t \right) &= D_{M} \frac{1}{s} \partial_{s} \left( s \partial_{s} \right) m \left( s,t \right) \nonumber \\*
	&\quad- k_{MB} m\left( s,t \right) + u_{BM} b\left(t\right), \label{eq:time_evol_m} \\
	V \partial_{t} b \left(t \right) &=
	\int_{0}^{L} \left[ k_{FB} f \left( z,t \right) - 2 \pi r u_{BF} b \left( t \right) \right] \diff{z} \nonumber \\*
	&\quad+ \int_{r}^{R} 2 \pi s \left[ k_{MB} m\left( s,t \right) - u_{BM} b\left(t\right) \right] \diff{s}, \label{eq:time_evol_b}
\end{align}
where we define volume of the bulk $V$,
\begin{equation}
	V = \pi \left( R^{2}-r^{2} \right) L.
	\label{eq:volum_def}
\end{equation}
Here, the filament density $f\!\left(z,t\right)$ represents the protein concentration per unit length along the filament,
which is the density per unit surface area of the filament multiplied by the angular factor $2\pi r$.
Hence, we will use units $\si{\per\micro\metre}$ for $f$, $\si{\per\micro\metre\squared}$ for $m$,
and $\si{\per\cubic\micro\metre}$ for $b$.
The boundary conditions of the partial differential equations are set by
the conservation of the number of particles and by the transition rates between the filament and the membrane,
which provide relations for the fluxes on the filament and on the membrane at their edges,
\begin{align}
	&\left[ D_{F} \partial_{z} f \left( z,t \right) + v_{F} f \left( z,t \right) \right]_{z=0} \nonumber \\*
	&\qquad\qquad= u_{FM} f \left( 0,t \right) - 2 \pi r u_{MF} m \left( r,t \right), \label{eq:bc_f_0} \\
	&\left[ D_{F} \partial_{z} f \left( z,t \right) + v_{F} f \left( z,t \right) \right]_{z=L}
	= 0, \label{eq:bc_f_L} \\
	&\left[ D_{M}  2 \pi s \partial_{s}  m \left( s,t \right) \right]_{s=r} \nonumber \\*
	&\qquad\qquad= -u_{FM} f \left( 0,t \right) + 2 \pi r u_{MF} m \left( r,t \right), \label{eq:bc_m_r} \\
	&\left[ D_{M}  2 \pi s \partial_{s}  m \left( s,t \right) \right]_{s=R} = 0. \label{eq:bc_m_R}
\end{align}
The system is greatly simplified by studying it in steady state, setting all time derivatives
in Eqs.\,\ref{eq:time_evol_f}, \ref{eq:time_evol_m}, and \ref{eq:time_evol_b} to zero and eliminating $t$ as a variable.
In steady state, integrating over Eqs.\,\ref{eq:time_evol_f} and \ref{eq:time_evol_m} and applying the boundary conditions
shows that Eq.\,\ref{eq:time_evol_b} becomes linearly dependent on Eqs.\,\ref{eq:time_evol_f} and Eq.\,\ref{eq:time_evol_m}.
The linear dependence is a consequence of the conservation of particles,
which imposes that steady state fluxes have to loop back to their origin.
Hence, we omit Eq.\,\ref{eq:time_evol_b} from the steady state equations.
Additionally, it is helpful to make the equations non-dimensional by defining the following dimensionless variables,
\begin{equation}
	\begin{array}{lll}
		\alpha = \frac{v_{F}}{\sqrt{D_{F} k_{FB}}}, &
		\beta = \frac{u_{FM}}{\sqrt{D_{F} k_{FB}}}, &
		\gamma = \frac{u_{MF}}{\sqrt{D_{M} k_{MB}}}, \\
		\delta = \frac{u_{MF} k_{FB} u_{BM}}{u_{FM} u_{BF} k_{MB}}, &
		\lambda = \sqrt{\frac{k_{FB}}{D_{F}}} z, &
		\Lambda = \sqrt{\frac{k_{FB}}{D_{F}}} L, \\
		\sigma = \sqrt{\frac{k_{MB}}{D_{M}}} s, &
		\rho = \sqrt{\frac{k_{MB}}{D_{M}}} r, &
		\Rho = \sqrt{\frac{k_{MB}}{D_{M}}} R.
	\end{array}
	\label{eq:nondimensionalisation_parameters}
\end{equation}
We also rescale the density fields to make them dimensionless,
\begin{align}
	\varphi \left( \lambda \right) &= \frac{k_{FB}}{2\pi r u_{BF} b} f \left(\sqrt{ \frac{D_{F}}{k_{FB}}} \lambda \right), \nonumber \\*
	\mu \left( \sigma \right) &= \frac{k_{MB}}{u_{BM} b} m \left( \sqrt{\frac{D_{M}}{k_{MB}}} \sigma \right).
	\label{eq:nondimensionalisation_fields}
\end{align}
Using these definitions and the steady state condition, Eq.\,\ref{eq:time_evol_f} and Eq.\,\ref{eq:time_evol_m} become
\begin{align}
	\partial_{\lambda}^{2}\varphi \left( \lambda \right) + \alpha \partial_{\lambda}\varphi \left( \lambda \right)
	&-\varphi \left( \lambda \right) + 1 = 0, \label{eq:ss_phi_dif} \\
	\frac{1}{\sigma} \partial_{\sigma} \left( \sigma \partial_{\sigma} \right) \mu \left( \sigma \right)
	&- \mu \left( \sigma \right) + 1 = 0. \label{eq:ss_mu_dif}
\end{align}
Using the dimensionless variables, the boundary conditions become
\begin{align}
	\left[ \partial_{\lambda} \varphi \left( \lambda \right) \right]_{\lambda = 0}
	&+ \alpha \varphi \left( 0 \right) = \beta \left( \varphi \left( 0 \right)
	- \delta \mu \left( \rho \right) \right), \label{eq:ss_bc_phi_0} \\
	\left[ \partial_{\lambda} \varphi \left( \lambda \right) \right]_{\lambda = \Lambda}
	&+ \alpha \varphi \left( \Lambda \right)  = 0, \label{eq:ss_bc_phi_Lambda} \\
	\left[\partial_{\sigma} \mu \left( \sigma \right) \right]_{\sigma = \rho}
	&= -\frac{\gamma}{\delta} \left( \varphi \left( 0 \right) - \delta \mu \left( \rho \right) \right), \label{eq:ss_bc_mu_rho} \\
	\left[\partial_{\sigma} \mu \left( \sigma \right) \right]_{\sigma = \Rho}& = 0. \label{eq:ss_bc_mu_Rho}
\end{align}
The general solutions of the ordinary differential equations Eq.\,\ref{eq:ss_phi_dif} and Eq.\,\ref{eq:ss_mu_dif}
are found by solving the homogeneous equations,
and adding the particular solutions $\varphi\!\left(\lambda\right)=1$ and $\mu\!\left(\sigma\right)=1$.
The full solutions read
\begin{align}
	\varphi \left( \lambda \right) = 1 + &C_{1} \exp \left[-\frac{\lambda}{2} \left( \sqrt{4+\alpha^{2}}+\alpha \right)\right] \nonumber \\*
	+ &C_{2} \exp \left[\frac{\lambda}{2} \left( \sqrt{4+\alpha^{2}}-\alpha \right)\right], \label{eq:phi_general_sol} \\
	\mu \left( \sigma \right) = 1 + &C_{3} K_{0} \left( \sigma \right) + C_{4} I_{0} \left( \sigma \right). \label{eq:mu_general_sol}
\end{align}
Here, $I_{0} \left( \sigma \right)$ and $K_{0} \left( \sigma \right)$ are
the modified Bessel functions of the first and second kind, respectively.
The integration constants $C_{1}$, $C_{2}$, $C_{3}$, and $C_{4}$ are determined by Eq.\ref{eq:ss_bc_phi_0}--\ref{eq:ss_bc_mu_Rho},
and are listed in Sec.\,\ref{sec:polarisation_solution_integration_constants}.
We can simply recover the protein density fields in the dimensionful representation  by using
the dimensionless exact solutions and substituting back the expressions listed in Eq.\,\ref{eq:nondimensionalisation_parameters},
providing exact solutions for the protein densities and for the protein flux in the case of $D_{B}\rightarrow \infty$.

Instead of the fourteen parameters of the full model,
the dimensionless model with fast bulk diffusion only contains seven independent parameters.
Three of those parameters set the system size ($\rho$, $\Rho$, and $\Lambda$),
and the four remaining independent parameters ($\alpha$, $\beta$, $\gamma$, and $\delta$) set the system dynamics.
Of these, $\beta$ and $\gamma$ determine how fast the particles transition between the membrane and the filament.

The parameters $\alpha$ and $\delta$ describe the non-equilibrium nature of the system.
When $\alpha=0$ and $\delta=1$, we find that $C_{1}=C_{2}=C_{3}=C_{4}=0$,
and the particle densities reach their equilibrium values
\begin{align}
	\varphi_{\mathrm{eq}} \left( \lambda \right) = 1,
	\label{eq:equil_phi} \\
	\mu_{\mathrm{eq}} \left( \sigma \right) = 1.
	\label{eq:equil_mu}
\end{align}
When the system is out of equilibrium, a steady-state flux can exist that on average brings particles from the bulk to the filament,
from the filament to the membrane, and from the membrane back to the bulk.
Eqs.\,\ref{eq:ss_bc_phi_0}--\ref{eq:ss_bc_mu_rho} show that this flux is proportional to
\begin{equation}
	J_{ss} \propto \varphi \left( 0 \right) - \delta \, \mu \left( \rho \right).
	\label{eq:steady_state_flux_polarisation}
\end{equation}
Using the exact solution, it can be shown that this flux vanishes if and only if $\alpha=0$ and $\delta=1$,
so these parameters determine whether detailed balance holds.

Because $\alpha$ and $\delta$ can be varied independently,
the model contains two essential processes by which detailed balance can be broken.
Firstly, a positive value of $\alpha$ represents a drift velocity on the filament, as shown in Eq.\,\ref{eq:nondimensionalisation_parameters}.
Here, motor proteins on the filament drive the particles and create a flux through the system,
where particles bind from the bulk to the filament, are driven towards the membrane to which they bind,
diffuse on the membrane moving away from the filament, and finally fall off into the bulk again.
Secondly, the value of $\delta$, defined in Eq.\,\ref{eq:nondimensionalisation_parameters},
describes the extent to which membrane and filament binding are in or out of equilibrium.
The particle itself can undergo a chemical modification step along the cycle filament-membrane-bulk-filament,
which is driven by the dissipation of chemical free energy and leads to a value $\delta<1$.
For example, a protein may exist in several phosphorylation states,
such as fission yeast Pom1 which has a high affinity for the the membrane when it is dephosphorylated at microtubule tips,
but quickly unbinds from the membrane when it is rephosphorylated there~\cite{hachet_phosphorylation_2011}.
If an NTP molecule is hydrolyzed at any step in the forward direction of the filament-membrane-bulk-filament cycle,
the model shows that a membrane spot is formed independent of whether particles are actively transported along the filament.
There are thus two distinct mechanisms that can act independently to create a polarized distribution of membrane proteins.

\begin{figure}
	\includegraphics{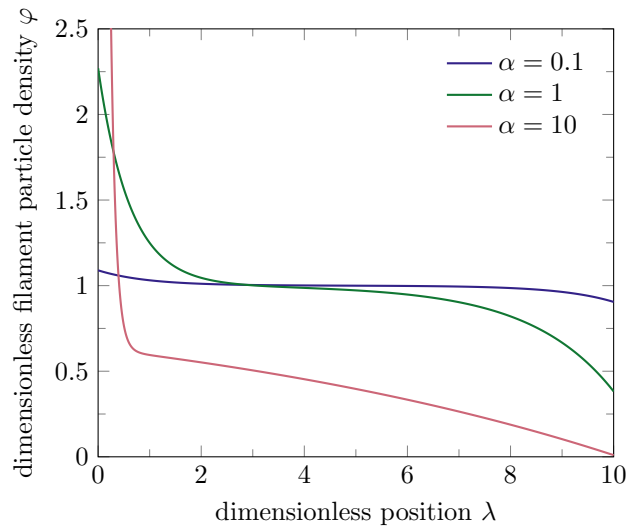}
	\caption{\label{fig:polarisation_phi_vs_lambda_vary_alpha}
		The dimensionless particle density along the filament $\varphi\!\left(\lambda\right)$ is shaped by
		the dimensionless drift velocity $\alpha = v_{F}/\sqrt{D_{F}k_{FB}}$.
		The driving force moves particles away from the right border at $\lambda=\Lambda$ towards the membrane at $\lambda=0$,
		lowering the particle density close to $\lambda=\Lambda$.
		Due to this reduced density on the right, the binding of new particles exceeds the unbinding as the particles move to the left,
		and the binding and unbinding balance out after a length scale $\lambda_{\mathrm{binding}}$.
		On the opposite side, particles are crowded against the membrane, which leads to a peak in the distribution with a size of $\lambda_{\mathrm{crowding}}$.
		For low drift velocities ($\alpha=0.1$), the distribution is close to the equilibrium shape $\varphi\!\left(\lambda\right)=1$,
		but for larger values ($\alpha=1$) the shape becomes more pronounced.
		When the drift velocity increases even more ($\alpha=10$), the amplitudes of the density deformations increase further,
		but are now accompanied by a decrease of the length scale $\lambda_{\mathrm{crowding}}$
		and an increase of the length scale $\lambda_{\mathrm{binding}}$.
		For $\alpha \approx \Lambda$, the full filament acts as an antenna for the adsorption of particles from the bulk,
		and roughly all those particles reach the membrane.
		If $\alpha<\Lambda$, many particles fall off the filament before they reach the membrane.
		We use the parameter values listed in Table~\ref{tab:polarisation_dimensionless_parameter_values},
		except for $\alpha$ and $\Lambda=10$.
	}
\end{figure}
The parameter $\alpha$ describes how fast the particles are driven on the filament,
\begin{equation}
	\alpha = \frac{v_{F}}{k_{FB}} / \sqrt{\frac{D_{F}}{k_{FB}}} = l_{v} / l_{D}.
	\label{eq:alpha_explanation}
\end{equation}
Here, $l_{v}$ is the average distance that a particle travels on the filament before it unbinds when it moves with a drift velocity $v_{F}$,
and $l_{D}$ sets a length scale over which diffusion smooths out the profile of the particle density on the filament $f\!\left(z\right)$.
Hence, $\alpha$ measures how much the particle density is shaped by the drift velocity.
In Fig.\,\ref{fig:polarisation_phi_vs_lambda_vary_alpha}, we show that the particle distribution on the filament
is reshaped more strongly when $\alpha$ increases.
Particles are transported from the right to the left, where the membrane is located at $\lambda=0$.
In the center of the filament, the rate at which particles bind from the bulk equals the rate at which particles unbind from the filament,
and transport of particles coming in from the right equals the transport out to the left.
Hence, the particle density is simply the equilibrium density there.
However, within a distance $\lambda_{\mathrm{binding}}$ from the end of the filament ($\lambda=\Lambda$),
the particle density decreases because no particles can be transported from beyond $\Lambda$
while particles are still transported towards the left.
Similarly, within a distance $\lambda_{\mathrm{crowding}}$ from the membrane ($\lambda=0$),
the particle density peaks because transport brings in particles from the right while they cannot be transported further to the left.
\begin{figure}
	\includegraphics{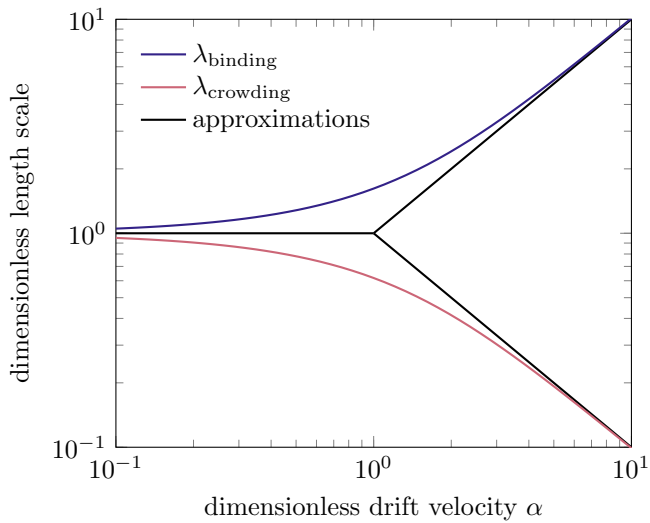}
	\caption{\label{fig:polarisation_length_scales_alpha}
		The dimensionless length scales $\lambda_{\mathrm{binding}}$ and $\lambda_{\mathrm{crowding}}$
		are roughly equal for $\alpha<1$, but when the driving velocity increases ($\alpha>1$)
		the particles can be transported for longer distances, which increases $\lambda_{\mathrm{binding}} \approx \alpha$.
		Furthermore, particles are pushed against the membrane more strongly, decreasing $\lambda_{\mathrm{crowding}} \approx 1/\alpha$.
		The exact expressions are given in Eq.\,\ref{eq:length_binding} and Eq.\,\ref{eq:length_crowding},
		while the approximations are given in Eqs.\,\ref{eq:small_alpha_approx_binding}--\ref{eq:large_alpha_approx_crowding}.
	}
\end{figure}
As shown in Eq.\,\ref{eq:phi_general_sol}, the two dimensionless length scales are given by
\begin{align}
	\lambda_{\mathrm{binding}} = \frac{2}{\sqrt{4+\alpha^{2}}-\alpha},
	\label{eq:length_binding} \\
	\lambda_{\mathrm{crowding}} = \frac{2}{\sqrt{4+\alpha^{2}}+\alpha}.
	\label{eq:length_crowding}
\end{align}
These length scales can be approximated in both the small and large limits of $\alpha$, showing that up to leading order
\begin{align}
	\lambda_{\mathrm{binding}} &\xrightarrow{\alpha \ll 1} 1,
	\label{eq:small_alpha_approx_binding} \\
	\lambda_{\mathrm{crowding}} &\xrightarrow{\alpha \ll 1} 1,
	\label{eq:small_alpha_approx_crowding} \\
	\lambda_{\mathrm{binding}} &\xrightarrow{\alpha \gg 1} \alpha,
	\label{eq:large_alpha_approx_binding} \\
	\lambda_{\mathrm{crowding}} &\xrightarrow{\alpha \gg 1} \frac{1}{\alpha}.
	\label{eq:large_alpha_approx_crowding}
\end{align}
These approximations are plotted in Fig.\,\ref{fig:polarisation_length_scales_alpha}
along with the exact length scales from Eq.\,\ref{eq:length_binding} and Eq.\,\ref{eq:length_crowding}.
The figure shows that for $\alpha<1$, the length scales $\lambda_{\mathrm{binding}}$ and $\lambda_{\mathrm{crowding}}$
barely change with $\alpha$, and increasing the driving velocity only increases the absolute slopes of the particle densities
at both ends of the filament, as seen in Fig.\,\ref{fig:polarisation_phi_vs_lambda_vary_alpha}.
But when $\alpha>1$, the larger drift velocity crowds the proteins tighter against the membrane,
creating a peak with a small $\lambda_{\mathrm{crowding}}$ close to $\lambda=0$.
Furthermore, the slope at the back of the filament becomes longer as the average distance that particles travel before they unbind increases.
This is known as the antenna effect~\cite{varga_yeast_2006,varga_kinesin-8_2009},
since the microtubule acts as an antenna that transports particles over a distance
that equals the motor protein processivity length $l_{v}$ (see Eq.\,\ref{eq:alpha_explanation}).
Hence, protein transport to the membrane is the most efficient when the length of the filament
equals the antenna length, $\Lambda=\lambda_{\mathrm{binding}}$.
If the filament is shorter, the protein concentration at the membrane will decrease.
However, if the filament is longer, then many proteins will fall off the filament before they arrive at the membrane,
wasting the chemical energy that was spent on driving them forward.

\begin{figure}
	\includegraphics{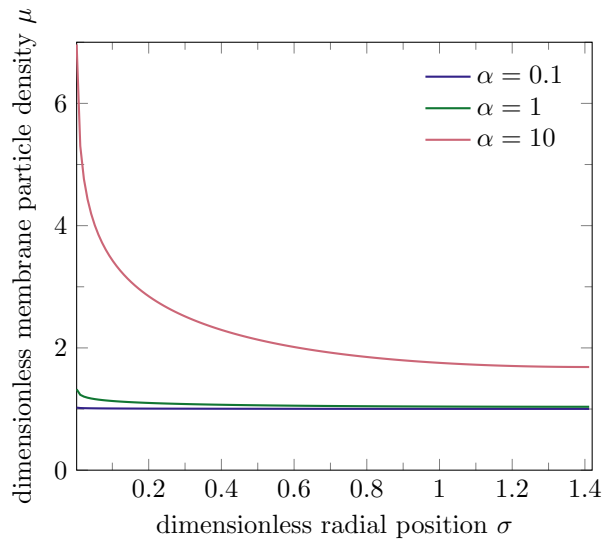}
	\caption{\label{fig:polarisation_mu_vs_sigma_vary_alpha}
		The dimensionless particle density on the membrane $\mu\!\left(\sigma\right)$ becomes more peaked
		when the dimensionless drift velocity $\alpha$ increases.
		There is a peak in the particle density on the filament as shown in Fig.\,\ref{fig:polarisation_phi_vs_lambda_vary_alpha},
		which is connected to the membrane at $\sigma=\rho\approx \num{1.8e-3}$.
		Some of these particles are deposited on the membrane, after which they diffuse away and finally unbind to the bulk.
		This diffusion and unbinding sets a length scale that equals $1$ in the dimensionless representation.
		If the particles are driven along the filament to the membrane faster (increasing $\alpha$),
		the height of the particle density on the membrane increases, but the width of the high density spot does not change.
		This width equals $\sqrt{D_{M}/k_{MB}}$, or \num{1} in the dimensionless system.
		The parameters are the same as in Fig.\,\ref{fig:polarisation_phi_vs_lambda_vary_alpha}.
	}
\end{figure}
Because the motor drift on the filament creates a high protein density on the filament end that is close to the membrane,
the interaction between the membrane and the filament will also lead to a higher protein density on the membrane.
As shown in Fig.\,\ref{fig:polarisation_mu_vs_sigma_vary_alpha},
this leads to the formation of a high density spot on the membrane close to the filament.
The density of the spot increases with $\alpha$, as the steady state particle flux around the cycle increases,
but the size of the spot $l_{s}$ is set by the unbinding rate and the diffusion constant,
\begin{equation}
	l_{s} = \sqrt{\frac{D_{M}}{k_{MB}}}.
	\label{eq:size_membrane_spot}
\end{equation}
Hence, the size of the spot is roughly constant, and the dimensionless density on the membrane close to the filament $\mu\!\left(\rho\right)$
fully measures how pronounced the membrane spot is.
If there is no spot, $\mu\!\left(\rho\right)=1$, and a value larger than unity shows how much higher the protein density is in the membrane spot
compared to the density on the membrane far away from the filament, which is set by the equilibrium dynamics between the bulk and the membrane.
Therefore, we will use $\mu\!\left(\rho\right)$ to assess the quality of polarization,
which allows us to measure the impact that different parameters have on cell polarization.

\begin{figure}
	\includegraphics{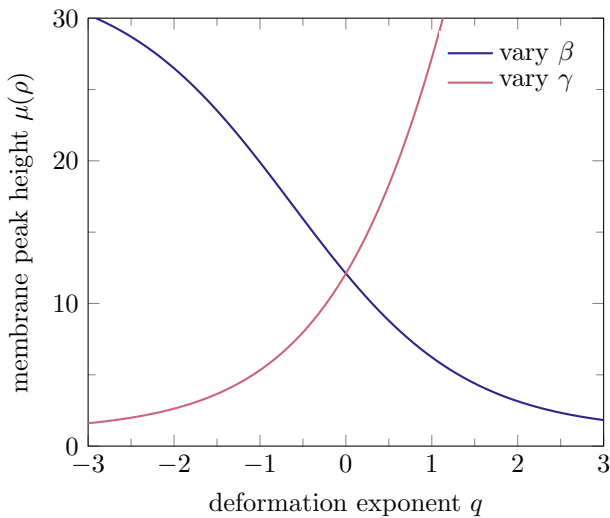}
	\caption{\label{fig:polarisation_mu_vs_beta_gamma_change}
		The measure for the height of the membrane spot, $\mu\!\left(\rho\right)$,
		changes when varying the dimensionless parameters $\beta$ and $\gamma$,
		which describe the rate of hopping from the filament to the membrane and vice versa, respectively.
		Here we vary these parameters and keep $\delta=1$, such that there is no free-energy drop
		associated with the binding and unbinding around the full cycle filament-membrane-bulk-filament,
		and detailed balance is only broken by active transport ($\alpha\approx11$).
		We vary either $\beta= \beta_{0} \exp\!\left(q\right)$ and $\gamma=\gamma_{0}$ (dark blue),
		or $\beta=\beta_{0}$ and $\gamma = \gamma_{0} \exp\!\left(q\right)$ (light red),
		where $\beta_{0}$ and $\gamma_{0}$ are the values shown in Table~\ref{tab:polarisation_dimensionless_parameter_values}.
		We see that if we increase the rate $\beta$ to move from the filament to the membrane
		while keeping $\delta=1$, the height of the spot decreases.
		Similarly, if we increase the rate $\gamma$ to move back from the membrane
		to the filament while keeping $\delta=1$, the polarization of the membrane distribution improves.
		Hence, when binding is in equilibrium, the affinity for the membrane should be low compared to the affinity for the microtubule,
		such that the particles that are deposited on the membrane by the microtubule have a large impact on the membrane distribution.
	}
\end{figure}
The dimensionless parameter $\beta$ quantifies how fast the particles move from the microtubule tip to the membrane,
and the parameter $\gamma$ quantifies the speed of the reverse transition.
In Fig.\,\ref{fig:polarisation_mu_vs_beta_gamma_change}, we show that the height of the membrane spot decreases
if we increase the rate of binding from the filament to the membrane or decrease the reverse rate, while keeping $\delta=1$.
This counter-intuitive result holds true precisely because $\delta=1$, such that
there is no net free-energy drop along the cycle filament-membrane-bulk-filament.
Increasing $\beta$ or decreasing $\gamma$ increases the affinity for the membrane relative to that for the filament,
but the constraint that there is no free-energy drop around the cycle means that
this must be accompanied by a decrease of the relative affinity for the filament compared to the bulk
or by an increase of the relative affinity for the membrane compared to the bulk.
In the former case, the density of particles at the tip of the filament decreases,
which reduces the height of the membrane spot because the flux from the filament tip to the membrane is decreased.
In the latter case, the equilibrium density on the membrane increases,
reducing the contrast between the membrane spot and the density far away.
The absolute equilibrium probability that the particle is found in the bulk
does not influence the relative likelihoods of finding the particle on the membrane or on the filament when the binding cycle is in equilibrium,
explaining why the details of the transition rates to and from the bulk become irrelevant in the non-dimensional representation.
Hence, increasing $\beta$ and decreasing $\gamma$ decreases $\mu\!\left(\rho\right)$
because the binding affinity for the filament should be high compared to the binding affinity for the membrane to create a polarized membrane spot.

\begin{figure}
	\includegraphics{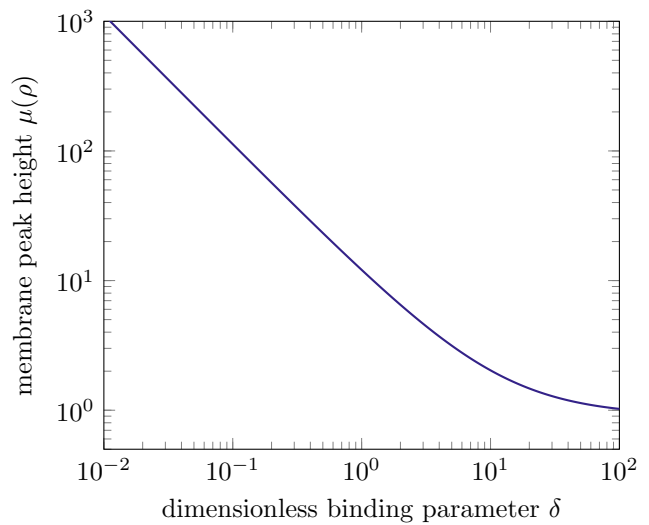}
	\caption{\label{fig:polarisation_mu_vs_delta_change}
		The measure for the height of the membrane spot relative to the equilibrium density on the membrane, $\mu\!\left(\rho\right)$,
		decreases when $\delta$ increases while the driving velocity remains constant ($\alpha\approx11$).
		To achieve a value $\delta \neq 0$, the binding cycle filament-membrane-bulk-filament needs to be coupled to the dissipation of chemical free energy.
		For example, performing an NTP hydrolysis each time the particle moves through the cycle leads to $\delta<1$,
		and the steady state flux through the cycle increases.
		Increasing this flux has a positive effect on the protein density in the membrane spot, and increases the polarization
		independent of the driving velocity on the filament $\alpha$.
		Similarly, creating an NTP molecule each cycle leads to $\delta>1$, and suppresses the flux and hence the polarization.
		Apart from $\delta$, the parameter values are given in Table~\ref{tab:polarisation_dimensionless_parameter_values}.
	}
\end{figure}
Finally, we focus on how $\delta$ affects the formation of a steady state membrane spot.
If $\delta<1$, then there is a free-energy drop each time a particle moves from
the bulk to the filament, from the filament to the membrane, and from the membrane back to the bulk.
The protein density in the membrane spot depends on the flux that runs through this cycle,
so we expect $\mu\!\left(\rho\right)$ to increase when $\delta$ decreases.
Fig.\,\ref{fig:polarisation_mu_vs_delta_change} shows that the height of the membrane spot indeed increases when $\delta<1$.
Furthermore, the shape of the density profile on the filament barely changes,
only slightly reducing the size of the density peak $\varphi\!\left(0\right)$ when $\delta<1$ (data not shown).
Hence, breaking detailed balance through a chemical modification cycle in the binding cycle filament-membrane-bulk-filament
has a strong effect on the formation of a polarized protein distribution on the membrane.

\section{\label{sec:parameter_values_membrane_polarisation}Biologically relevant parameter values}
The dimensionful model presented in Sec.\,\ref{sec:polarisation_model} contains fourteen parameters,
including one that sets the average concentration $b$.
To focus the model analysis, we find a set of biologically relevant parameter values,
which also provide values for the dimensionless parameter set.
Then, we vary individual parameters to investigate how they influence the polarization of the membrane and the free-energy dissipation.

Because microtubules are roughly \SI{25}{\nano\metre} in diameter~\cite{wade_microtubule_1997},
we choose a filament radius $r=\SI{12.5}{\nano\metre}$.
Furthermore, we choose a filament length of $L=\SI{10}{\micro\metre}$, which is a typical microtubule length.
For the container radius, we choose a value of $R=\SI{10.0125}{\micro\metre}$,
which will be large enough to show the full membrane spots that are created by a single microtubule,
and which ensures that $R-r=\SI{10}{\micro\metre}$, simplifying the creation of histograms that cover the entire radial space of the membrane.
This leads to a container that is large compared to the typical size of a fission yeast cell,
which is roughly \SI{10}{\micro\metre} long~\cite{mitchison_growth_1985}.
However, the exact size of the spot, and thus the required container size, does not influence the conclusions of this chapter.
Furthermore, we will discuss in the following paragraphs that we expect that the diffusion constant of proteins on the membrane is effectively lower
in vivo compared to in vitro, creating a smaller spot that would fit on a yeast cell.

The dynamical parameters on the filament describe how a cargo moves on a microtubule under the influence of motor proteins.
Single kinesin motors can move between \SI{0.01}{micro\metre\per\second} and \SI{1}{\micro\metre\per\second} depending on the ATP concentration,
the load, and the kind of kinesin protein~\cite{schnitzer_force_2000,guo_force_2019-1}.
Since we consider low load single protein cargoes, we set the average drift velocity to $v_{F} = \SI{0.5}{\micro\metre\per\second}$.
The dynamics on the filament is defined by this drift velocity in combination with the diffusion constant of kinesin motors running along a microtubule,
which has been measured to be roughly between \SI{0.002}{\micro\metre\squared\per\second}~\cite{okada_processive_1999}
and \SI{0.005}{\micro\metre\squared\per\second}~\cite{bormuth_breaking_2010}.
Hence, we set $D_{F}=\SI{0.004}{\micro\metre\squared\per\second}$.

The diffusion constant on the membrane depends on the diffusing protein and the type of lipid bilayer.
For a simple membrane in vitro, the diffusion constant of proteins with a radius of roughly \SI{2}{\nano\metre}
equals around \SI{8}{\micro\metre\squared\per\second}~\cite{weis_quantifying_2013}.
Because we employ a minimal model, we ignore the compartmentalization of the membrane~\cite{fujiwara_phospholipids_2002},
and choose a diffusion constant on the membrane of $D_{M}=\SI{5}{\micro\metre\squared\per\second}$.
In the bulk, we require the diffusion constant of proteins in the cytosol,
which was measured to be on average $D_{B}=\SI{60}{\micro\metre\squared\per\second}$~\cite{kuhn_protein_2011}.

We assume that the rate at which the cargo unbinds from the microtubule is limited by the unbinding rate of kinesin,
which is roughly $k_{FB}=\SI{0.5}{\per\second}$ when no external force pulls on the motor~\cite{guo_force_2019}.
For the rate at which the particles unbind from the membrane, we use the experimental off-rate
that was measured for the Rho-GTPase Cdc42, which sets a value of $k_{MB}=\SI{0.1}{\per\second}$~\cite{johnson_new_2009}.
Together with the membrane diffusion constant $D_{M}$, this rate sets a typical length scale for the membrane spot size of
\begin{equation}
	l_{M} = \sqrt{\frac{D_{M}}{k_{MB}}} \approx \SI{7}{\micro\metre}.
	\label{eq:membrane_spot_size}
\end{equation}
The same experiments also provide an order of magnitude for the binding rate of particles from the bulk to the membrane
by observing the Cdc42 association with liposomes.
In equilibrium, the total binding rate from the solution equals
\begin{equation}
	k_{\mathrm{on}} = u_{BM} A b,
	\label{eq:total_binding_rate_to_membrane}
\end{equation}
where $A$ is the area of the membrane on spherical liposomes.
Taking a concentration of $b=\SI{50}{\nano M}$, a liposome radius of \SI{0.5}{\micro\metre} (giving $A\approx\SI{3.1}{\micro\metre\squared}$),
and a $k_{\mathrm{on}}=\SI{1}{\per\second}$~\cite{johnson_new_2009}, we find $u_{BM} \approx \SI{0.01}{\micro\metre\per\second}$.

The binding rate of cargo proteins from the cytosol to the filament is affected by the binding rate
of motor proteins to the microtubule and the binding rate of the cargo to motor proteins.
We assume that this binding is relatively strong, such that the equilibrium affinity for the microtubule is much higher than for the membrane,
which allows the driving on the microtubule to have a pronounced effect on the membrane concentration.
Because the rate of unbinding from the filament to the bulk is larger than that from the membrane to the bulk, $k_{FB} = 5 k_{MB}$,
this means that $u_{BF}$ must be several orders of magnitude larger than $u_{BM}$
to guarantee a higher equilibrium affinity for the filament.
Furthermore, because the area of the microtubule is much smaller than the area of the membrane,
the affinity for the microtubule needs to be very large to find a significant fraction of particles on the filament.
Hence, we choose $u_{BF} = \SI{100}{\micro\metre\per\second}$.
We can compare our values of $k_{FB}$ and $u_{BF}$ to the reported dissociation constant $K_{D,BF}$
of a kinesin subunit binding to microtubules, which was found to be lower than
$K_{D,BF}<\SI{50}{\nano M} \approx \SI{30}{\per\micro\metre\cubed}$~\cite{hackney_implications_1995}.
This dissociation constant should equal
\begin{equation}
	K_{D,BF} = \frac{k_{FB}}{u_{BF}A_{\mathrm{tub}}},
	\label{eq:dissociation_constant_kinesin_MT}
\end{equation}
where $A_{\mathrm{tub}}$ is the surface area of a single tubulin dimer on the microtubule, which we take to be roughly \SI{50}{\nano\metre\squared}.
Using the previously mentioned values of $k_{FB}$ and $u_{BF}$, we find $K_{D,BF}\approx \SI{100}{\per\micro\metre\cubed}$.
Hence, the experimental value of the dissociation constant suggests that we slightly underestimate the binding to the microtubule,
but our value is of the right order of magnitude.

Once the particles are driven to the tip of the filament, and are in contact with the membrane,
the rate to bind to the membrane $u_{FM}$ is related to the rate $u_{BM}$.
If the mechanism of binding from the filament is the same as from the bulk,
then we assume $u_{FM}=u_{BM}=\SI{0.01}{\micro\metre\per\second}$.
However, the cell may implement a different mechanism for transporting the cargo from the tip of a microtubule to the membrane,
in which case this value could be larger.
Hence, it will be interesting to see how the prominence of the membrane spot changes with $u_{FM}$.

Finally, detailed balance provides a relation for the reverse rate $u_{MF}$ at which particles
that are on the membrane and close to the filament bind to the filament.
If the binding and unbinding is in equilibrium, we have
\begin{equation}
	\delta = \frac{u_{MF} k_{FB} u_{BM}}{u_{FM} u_{BF} k_{MB}} = 1.
	\label{eq:detailed_balance_polarisation_relation_for_uMF}
\end{equation}
Using the previously mentioned values for all other (un)binding rates, we find $u_{MF} = \SI{1}{\micro\metre\per\second}$.
In Table~\ref{tab:polarisation_simulation_parameter_values}, we give an overview of all parameter values,
and in Table~\ref{tab:polarisation_dimensionless_parameter_values} we list the values
that the dimensionless parameters take in the biologically relevant regime.

\section{\label{sec:polarisation_simulations}A finite bulk diffusion constant is beneficial for cell polarization}
In steady state, Eqs.\,\ref{eq:time_evol_f}--\ref{eq:time_evol_b} provide a set of ordinary differential equations
that can be solved analytically, as shown in Eq.\,\ref{eq:phi_general_sol}--\ref{eq:mu_general_sol}.
However, these equations are valid when the diffusion constant in the bulk is very large, $D_{B}\rightarrow \infty$.
When $D_{B}$ is finite, we could modify the steady state differential equations and the corresponding boundary conditions,
leading to a set of coupled partial differential equations for $f\!\left(z\right)$, $m\!\left(s\right)$ and $b\!\left(z,s\right)$
that cannot be solved analytically.
To find numerical solutions for the steady state density profiles, we perform simple time step based Monte Carlo simulations
of a single particle moving through the system.
We do not require multiple particles since we model an "ideal gas" of proteins that never interact.

Because of the radial symmetry of the system, we only keep track of the longitudinal position $z$ and the radial distance $s$.
To model the particle diffusion, we implement a discrete time random walk in which particles move by
a stochastic longitudinal step $\delta z$ or radial step $\delta s$ each time step $\delta t$.
On the filament, $\delta z$ follows a Gaussian distribution with mean $-v_{F} \delta t$ and standard deviation $\sqrt{2 D_{F} \delta t}$,
while in the bulk the mean of $\delta z$ vanishes and the standard deviation equals $\sqrt{2 D_{B} \delta t}$.
For the diffusion in the radial direction, we draw two Gaussian random numbers $\delta x$ and $\delta y$
with zero mean and standard deviation $\sqrt{2 D_{M} \delta t}$ (membrane) or $\sqrt{2 D_{B} \delta t}$ (bulk).
Then, we calculate $(s+\delta s)^{2} = (s+\delta x)^{2} + \delta y^{2}$, where we use that the $x$-axis can always be chosen to point in the radial direction.
When either $\delta z$ or $\delta s$ brings the particle outside of the container, we reflect the particle position back across the boundary.
The algorithm to reflect the particles in a flat or circular surface without breaking detailed balance is described in Sec.\,\ref{sec:reflection_algorithms}.

To simulate the particle transitions between the three system parts,
we use the same kinetic Monte Carlo algorithm that we used previously~\cite{wierenga_diffusible_2020}.
In summary, we integrate the transition rates until a stochastically chosen threshold is passed
upon which we perform a reaction~\cite{prados_dynamical_1997}.
Since the binding rates $u_{BF}$, $u_{FM}$, $u_{MF}$, and $u_{BM}$ contain a spatial dimension as well,
we define the reaction length scales $l_{\mathrm{long}}$ and $l_{\mathrm{rad}}$.
When a particle is in the bulk and within a distance $l_{\mathrm{long}}$ from the membrane,
it can bind to the membrane with a transition rate
\begin{equation}
	k_{BM} = \frac{u_{BM}}{l_{\mathrm{long}}}.
	\label{eq:transition_rate_definition_bulk_membrane}
\end{equation}
This definition ensures that if the bulk concentration is roughly constant within the reaction volume,
then the flux is modeled correctly, $k_{BM} l_{\mathrm{long}} b \approx u_{BM} b$.
Hence, this equation is only valid when $l_{\mathrm{long}}$ is much smaller than the size of typical density fluctuations.
Similarly, we have
\begin{equation}
	k_{FM} = \frac{u_{FM}}{l_{\mathrm{long}}}.
	\label{eq:transition_rate_definition_filament_membrane}
\end{equation}
In the radial direction close to the filament, the reaction volume has a tubular form,
and we must have that $k_{BF} \pi \left(\left(r+l_{\mathrm{rad}}\right)^{2} - r^{2}\right) b \approx 2 \pi r u_{BF} b$.
Therefore, we choose
\begin{equation}
	k_{BF} = \frac{u_{BF}}{l_{\mathrm{rad}} + l_{\mathrm{rad}}^{2}/2r}
	\label{eq:transition_rate_definition_bulk_filament}
\end{equation}
and
\begin{equation}
	k_{MF} = \frac{u_{MF}}{l_{\mathrm{rad}} + l_{\mathrm{rad}}^{2}/2r}.
	\label{eq:transition_rate_definition_membrane_filament}
\end{equation}
When a reverse reaction occurs, we place the particle uniformly within the reaction volume.
For example, when a particle unbinds from the filament to the bulk,
it will be placed at a radial position $r<s<r+l_{\mathrm{rad}}$,
which is uniformly distributed over the area $\pi \left(\left(r+l_{\mathrm{rad}}\right)^{2} - r^{2}\right)$.
This mechanism ensures that detailed balance holds when $\alpha=0$ and $\delta=1$.

\begin{figure}
	\includegraphics{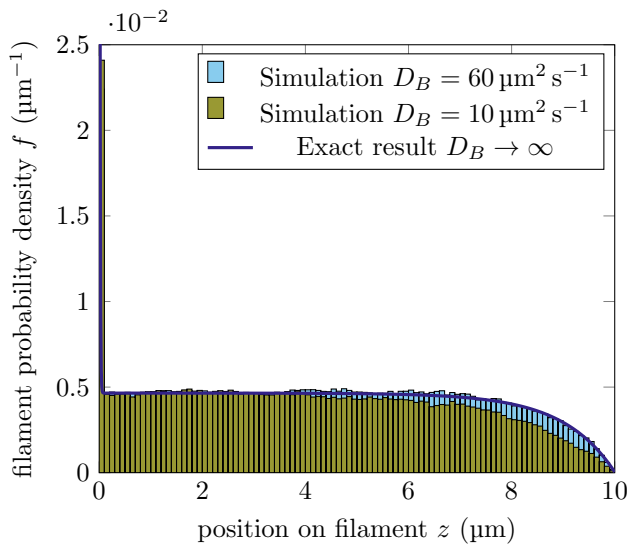}
	\caption{\label{fig:polarisation_filament_histogram}
		The probability density $f\!\left(z\right)$ that a particle is at position $z$ on the filament,
		where the membrane is located at $z=0$ and motor proteins drive the particles from the right to the left.
		We use the biologically relevant parameters of Table~\ref{tab:polarisation_parameter_values}.
		The dark blue line gives the exact result for $D_{B}\rightarrow \infty$, showing that the filament acts as an antenna
		over a length $l_{\mathrm{binding}} \approx \SI{1.0}{\micro\metre}$.
		After this distance, the unbinding of particles balances out the binding of new particles.
		When the particles reach the membrane, they are crowded over a length scale $l_{\mathrm{crowding}} \approx \SI{8}{\nano\metre}$.
		In light blue, we show a histogram of the particle density in simulations using the parameters of
		Table~\ref{tab:polarisation_simulation_parameter_values}, in which the bulk diffusion constant $D_{B}=\SI{60}{\micro\metre\squared\per\second}$,
		which nearly perfectly coincides with the theoretical result.
		Simulations with a lower bulk diffusion constant $D_{B}=\SI{10}{\micro\metre\squared\per\second}$ are shown in yellow,
		revealing that the antenna effect acts over a longer length scale when the bulk diffusion constant is lower.
		The lower diffusion constant allows some particles that unbind from the filament to rapidly rebind to the filament,
		increasing the effective distance over which the particles can be transported.
		The density peak in the leftmost bin is predicted to be \SI{2.1e-2}{\per\micro\metre} ($D_{B}\rightarrow\infty$),
		and the simulations find \SI{2.2e-2}{\per\micro\metre} ($D_{B}=\SI{60}{\micro\metre\squared\per\second}$)
		and \SI{2.4e-2}{\per\micro\metre} ($D_{B}=\SI{10}{\micro\metre\squared\per\second}$),
		indicating that the particle density at the filament tip increases when the bulk diffusion constant decreases.
	}
\end{figure}
\begin{figure}
	\includegraphics{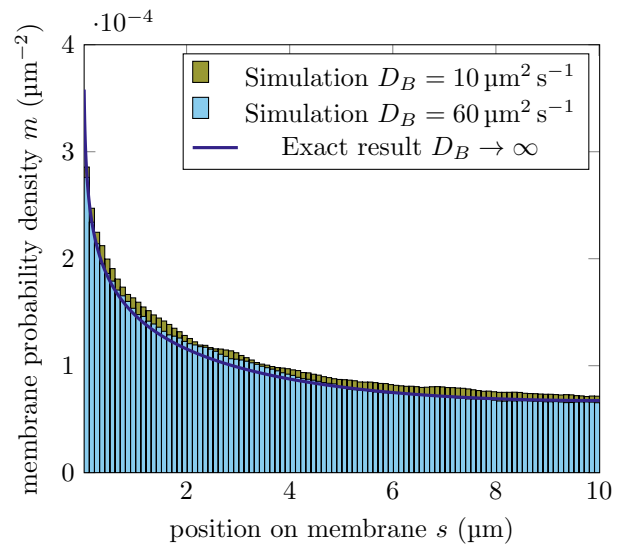}
	\caption{\label{fig:polarisation_membrane_histogram}
		The probability density $m\!\left(s\right)$ that a particle is at a radial position $s$ on the membrane,
		where the filament is located at $s=r=\SI{0.0125}{\micro\metre}$.
		We use the same parameters and simulation as in Fig.\,\ref{fig:polarisation_filament_histogram}.
		The dark blue line gives the exact result for $D_{B}\rightarrow \infty$,
		which predicts a spot size of $\sqrt{D_{M}/k_{MB}} = \SI{7.1}{\micro\metre}$.
		The difference between the simulations with a high bulk diffusion constant ($D_{B}=\SI{60}{\micro\metre\squared\per\second}$)
		and the exact solution are likely caused by insufficient sampling.
		The particle density in the leftmost bin is predicted to be \SI{2.7e-4}{\per\micro\metre\squared} ($D_{B}\rightarrow\infty$),
		and the simulations find \SI{2.8e-4}{\per\micro\metre\squared} ($D_{B}=\SI{60}{\micro\metre\squared\per\second}$)
		and \SI{2.9e-4}{\per\micro\metre} ($D_{B}=\SI{10}{\micro\metre\squared\per\second}$).
		The higher density for the lower bulk diffusion constant persists inside the whole membrane spot (yellow bins),
		showing that a lower bulk diffusion constant can slightly improve cell polarization.
	}
\end{figure}
We show the simulated protein concentration profiles on the filament and on the membrane in
Fig.\,\ref{fig:polarisation_filament_histogram} and Fig.\,\ref{fig:polarisation_membrane_histogram}, respectively,
and compare them to the exact solutions for $f\!\left(z\right)$ and $m\!\left(s\right)$ that are valid at $D_{B}\rightarrow \infty$.
We use the biologically relevant parameters listed in Table~\ref{tab:polarisation_simulation_parameter_values},
and the figures show that the simulations with $D_{B}=\SI{60}{\micro\metre\squared\per\second}$ agree with the analytical solutions.
In contrast, additional simulations with a bulk diffusion constant $D_{B}=\SI{10}{\micro\metre\squared\per\second}$ do display significant deviations
from the predicted behavior for $D_{B}\rightarrow \infty$.
Specifically, it appears that the length scale $l_{\mathrm{binding}}$ on the filament increases, improving the antenna effect caused by the motor proteins.
Furthermore, the density peak on the filament close to $z=0$ is higher when the bulk diffusion constant is lower.
This subsequently causes a larger flux of particles to be deposited on the membrane,
leading to a slight increase of the particle density in the membrane spot as shown in Fig.\,\ref{fig:polarisation_membrane_histogram}.

\begin{figure}
	\includegraphics{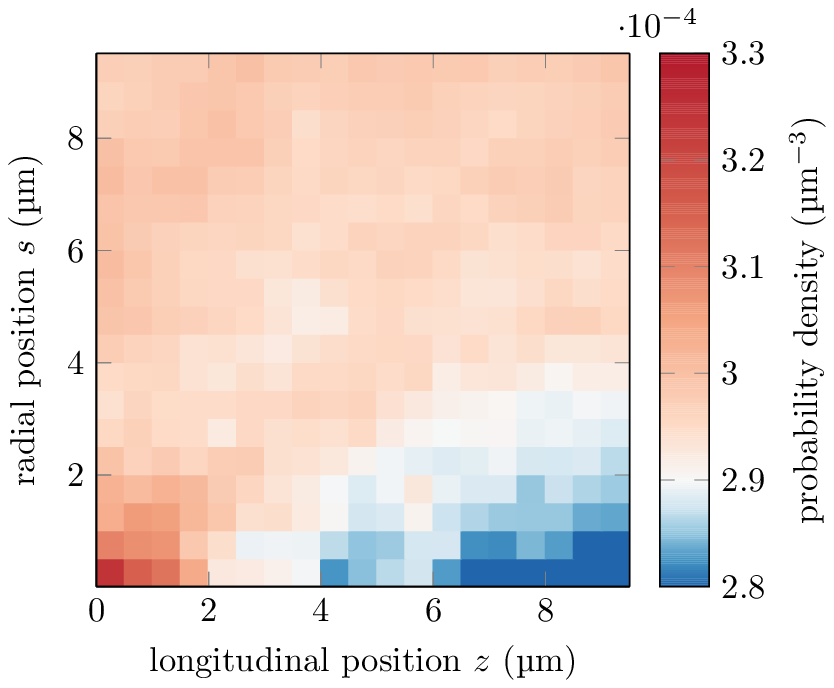}
	\caption{\label{fig:polarisation_bulk_histogram}
		The probability density $b\!\left(z,s\right)$ that a particle is at a longitudinal position $z$ and a radial position $s$ in the bulk,
		where the membrane is located at $z=0$ and the filament is located at $s=r=\SI{0.0125}{\micro\metre}$.
		We show the histogram produced by simulations with a lower diffusion constant in the bulk, $D_{B}=\SI{10}{\micro\metre\squared\per\second}$,
		because $D_{B}=\SI{60}{\micro\metre\squared\per\second}$ results in a nearly flat density profile.
		Otherwise, all parameters are the same as listed in Tab.\,\ref{tab:polarisation_parameter_values}.
		With a lower diffusion constant, the density profile on the filament shown in
		Fig.\,\ref{fig:polarisation_filament_histogram} persists for some distance into the bulk,
		improving the effects of active transport on the polarized particle distribution.
	}
\end{figure}
The steady state particle density in the bulk $b\!\left(z,s\right)$ is almost homogeneous when
$D_{B}=\SI{60}{\micro\metre\squared\per\second}$ (data not shown),
explaining why the exact solution for $D_{B}\rightarrow \infty$ is nearly identical to the simulation results.
To visualize the effects of a lower bulk diffusion constant,
we plot how the proteins are distributed in the cytosol for $D_{B}=\SI{10}{\micro\metre\squared\per\second}$ in Fig.\,\ref{fig:polarisation_bulk_histogram}.
Close to the filament, which is located at $s=r$, the bulk density profile resembles the shape of the density on the filament
shown in Fig.\,\ref{fig:polarisation_filament_histogram}.
The transport along the filament removes particles from the back of the container,
lowering the protein density in a region that extends several microns into the bulk.
Furthermore, Fig.\,\ref{fig:polarisation_bulk_histogram} shows that the particles are concentrated more densely
close to $z=0$ and $s=r$, which is close to the density peak on the filament and on the membrane.
Since a finite diffusion constant does not smear out the protein distributions instantly,
lowering the diffusion constant in the bulk is beneficial for the formation of a strongly polarized non-equilibrium distribution on the membrane.

\section{\label{sec:polarisation_free_energy}Free-energy is dissipated to form a membrane spot}
Using the exact solutions of the protein densities on the filament and the membrane in Eqs.\,\ref{eq:phi_general_sol}--\ref{eq:mu_general_sol},
we saw that the flux through the system vanishes if and only if $\delta=1$ and $\alpha=0$.
The parameter $\delta$ can break detailed balance when the particles undergo
a driven chemical modification cycle when they revolve through the filament, membrane, and bulk.
On the other hand, a non-zero value for the parameter $\alpha$ is caused by motor proteins
driving the movement of the particles along the filament.
Both processes dissipate free energy, and we can find lower bounds for the free-energy dissipation caused by each process.

On the filament, the particles have a drift velocity $v_{F}$ and a diffusion constant $D_{F}$.
Using the Einstein relation~\cite{einstein_uber_1905},
these two parameters define the average drift force $F_{F}$ that acts on particles bound to the filament,
\begin{equation}
	F_{F} = k_{B}T \frac{v_{F}}{D_{F}}.
	\label{eq:drift_force_from_einstein}
\end{equation}
We can calculate the work performed on a particle by multiplying this force with
the net distance that the particle travels on the filament.

For the binding reactions, we make use of the local detailed balance relation~\cite{maes_local_2020}.
Because the individual transitions can also involve the changing of dimensions, for example from the bulk to the membrane,
it is nonsensical to assign free-energy differences between particle states on the membrane or in the bulk,
\begin{equation}
	\Delta \mathcal{F}_{BM} \neq k_{B}T \log \!\left[\frac{u_{BM}}{k_{MB}}\right].
	\label{eq:local_detailed_balance_dimension_break}
\end{equation}
In this expression, we would take the logarithm of a factor that has a dimension of length.
Still, because a particle that moves through a loop has to pass the dimensional transitions along the $z$- and $s$-axis in both directions,
these length scales have to cancel in the ratio $\delta$.
Only a multiplicative constant $c$ may survive because the length scales of the transitions from the filament to the membrane
and from the bulk to the membrane,
or the length scales in the transitions from the membrane to the filament and from the bulk to the filament could differ.
Hence,
\begin{align}
	\Delta \mathcal{F}_{FMB} &= k_{B}T \log \! \left[ \frac{u_{FM} k_{MB} u_{BF} }{u_{MF} u_{BM} k_{FB}} \right]
	+ k_{B}T \log\left[c\right] \nonumber \\*
	&= -k_{B}T \log \! \left[ \delta \right] + k_{B}T \log\left[c\right].
	\label{eq:polarisation_local_detailed_balance_1}
\end{align}
Here, we take the dissipated free energy to be positive, such that a positive free-energy drop around the loop $\Delta \mathcal{F}_{FMB}$
drives a flux through the loop filament-membrane-bulk-filament.
Because we have shown that detailed balance holds if and only if $\delta=1$, and we require $\Delta \mathcal{F}_{FMB}=0$ in that case,
we see that the length scale $c=1$, showing that
\begin{equation}
	\Delta \mathcal{F}_{FMB} = -k_{B}T \log \! \left[ \delta \right].
	\label{eq:polarisation_local_detailed_balance_2}
\end{equation}

We can use Eq.\,\ref{eq:drift_force_from_einstein} and Eq.\,\ref{eq:polarisation_local_detailed_balance_2} to define free-energy differences,
but to compare the two we require the average dissipated power per particle.
In the simulations, we can keep track of the net distance traveled on the filament $d_{\mathrm{net}}$,
which increases if motor proteins drive the particle towards $z=0$, but decreases again if the diffusion with diffusion constant $D_{F}$ increases $z$.
Hence, we take into account that motor proteins that hydrolyze ATP have a small chance of moving back, reattaching a phosphate group to an ADP molecule.
Similarly, we record the net number of times the particle moves forward in the transition filament-membrane $N_{FM}$.
We run the simulations for a time $t$, after which the total average dissipated power $P_{\mathrm{sim}}$ equals
\begin{equation}
	P_{\mathrm{sim}} = \frac{ F_{F} d_{\mathrm{net}} + N_{FM} \Delta \mathcal{F}_{FMB}}{t}.
	\label{eq:dissipated_power_simulations}
\end{equation}
If the simulated time $t$ is long enough, we will find a reliable measurement of the power.

Using the steady state versions of Eq.\,\ref{eq:time_evol_f}--\ref{eq:bc_m_R} and their analytical solutions,
we can also calculate the exact dissipated power per particle for $D_{B}\rightarrow \infty$.
First, we see that the steady state flux that moves through the filament-membrane-bulk-filament cycle equals
\begin{equation}
	J_{ss} = u_{FM} f\!\left(0\right) - 2 \pi r u_{MF} m\!\left(r\right).
	\label{eq:polarisation_steady_state_flux}
\end{equation}
Then, the power dissipated by the binding dynamics equals
\begin{equation}
	P_{\mathrm{exact},FMB} = J_{ss} \Delta \mathcal{F}_{FMB}.
	\label{eq:dissipated_power_exact_biding}
\end{equation}
Then, the power dissipated by the driving on the filaments follows from the flux along the filament,
\begin{equation}
	P_{\mathrm{exact},F} = \int_{0}^{L} \left(D_{F} \partial_{z} f\!\left(z\right) + v_{F} f\!\left(z\right)\right) F_{F} \diff z,
	\label{eq:dissipated_power_exact_driving}
\end{equation}
where the diffusive term includes the free energy that is returned if a motor protein makes a backward step.
By integrating over the differential equations Eq.\,\ref{eq:time_evol_f}--\ref{eq:time_evol_b},
we can find expressions for the number of particles in each part of the system,
\begin{align}
	N_{F} &= \int_{0}^{L} f\!\left(z\right) \diff z \nonumber \\*
	&= \frac{1}{k_{FB}} \left(2\pi r L u_{BF} b - J_{ss}\right), \label{eq:number_of_particles_filament} \\
	N_{M} &= \int_{r}^{R} m\!\left(s\right) 2 \pi s \diff s \nonumber \\*
	&= \frac{1}{k_{MB}} \left( \pi \left(R^{2}-r^{2}\right) u_{BM} b + J_{ss} \right), \label{eq:number_of_particles_membrane} \\
	N_{B} &= \int_{0}^{L} \int_{r}^{R} b 2\pi s \diff s \diff z \nonumber \\*
	&= \pi \left(R^{2}-r^{2}\right) L b, \label{eq:number_of_particles_bulk} \\
	N_{\mathrm{tot}} &= N_{F}+N_{M}+N_{B}.
	\label{eq:number_of_particles_system_parts}
\end{align}
Using these definitions, we find the power per particle to equal
\begin{align}
	P_{\mathrm{exact}} &= \frac{D_{F} \left(f\!\left(L\right) - f\!\left(0\right)\right) + v_{F} N_{F}}{N_{\mathrm{tot}}} F_{F} \nonumber \\*
	&\qquad+ \frac{J_{ss}}{N_{\mathrm{tot}}} \Delta \mathcal{F}_{FMB}.
	\label{eq:dissipated_power_exact}
\end{align}
This solution is exact, but only valid when the diffusion constant in the bulk is very large, $D_{B}\rightarrow \infty$.

In Sec.\,\ref{sec:polarisation_simulations}, we showed that lowing the bulk diffusion constant slightly improves the quality of polarization,
but we also expect that it improves its efficiency.
Particles that are transported along the filament and fall off before they reach the membrane instantly diffuse away when $D_{B}\rightarrow\infty$.
Hence, the free energy that is dissipated in transporting these particles is wasted as soon as they fall off the filament.
However, a lower bulk diffusion constant allows the particles to rebind to the filament,
reducing the number of wastefully transported particles.
For the parameter values listed in Table~\ref{tab:polarisation_parameter_values},
we predict that the power of active transport equals $\SI{2.61}{k_{B}T\per\second}$ for $D_{B}\rightarrow \infty$,
and we find a power of $\SI{2.6}{k_{B}T\per\second}$ in simulations with $D_{B}=\SI{60}{\micro\metre\squared\per\second}$
and a power of $\SI{2.4}{k_{B}T\per\second}$ in simulations with $D_{B}=\SI{10}{\micro\metre\squared\per\second}$.
Hence, simulations show that lowering the bulk diffusion constant can indeed improve the efficiency of active transport
by increasing the quality of polarization and decreasing wasteful free-energy dissipation.

\section{\label{sec:polarisation_free_energy_costs}Driven binding kinetics is more efficient than motor transport}
Using the results of Sec.\,\ref{sec:polarisation_free_energy}, we can quantify
both the dissipated power by the motor proteins and by the binding kinetics of the cargo proteins.
Furthermore, we showed in Sec.\,\ref{sec:polarisation_model} that the quality of the polarized protein distribution on the membrane
can be quantified by $\mu\!\left(\rho\right)$, which compares the protein density in the center of the membrane spot
to the density on the periphery of the membrane.
Starting from the biologically relevant set of parameter values listed in Table~\ref{tab:polarisation_parameter_values},
we investigate how varying some parameters influences the membrane spot.

\begin{figure}
	\includegraphics{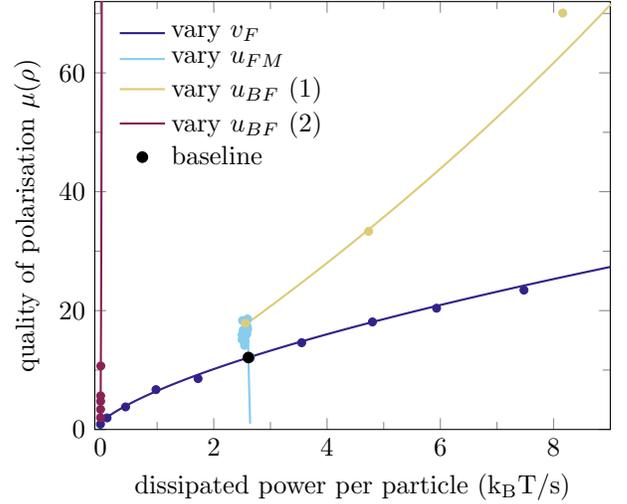}
	\caption{\label{fig:polarisation_vs_dissipation}
		The quality of polarization, given by the protein density at the center of the spot relative to
		the equilibrium density on the membrane, $\mu\!\left(\rho\right)$,
		as a function of the dissipated power in the system.
		The black point corresponds to the biologically relevant parameter values shown in Table~\ref{tab:polarisation_parameter_values},
		providing a baseline from where to change parameters.
		We vary the drift velocity on the filament $v_{F}$ between \num{0} and \SI{1}{\micro\metre\per\second},
		where the baseline value is given by $v_{F}=\SI{0.5}{\micro\metre\per\second}$,
		and we keep the binding in equilibrium (dark blue line, power law).
		We also vary the binding rate from the filament to the membrane $u_{FM}$ between \SI{7e-5}{\micro\metre\per\second}
		and \SI{1.5}{\micro\metre\per\second} (baseline $u_{FM}=\SI{0.01}{\micro\metre\per\second}$, light blue vertical line).
		This binding lowers the highly crowded density at the front of the filament, improving polarization while slightly saving on dissipated power.
		However, increasing $u_{FM}$ past $v_{F}$ does not influence the polarization.
		Hence, we keep the optimal value $u_{FM}=\SI{0.5}{\micro\metre\per\second}$ and increase the power dissipated through binding
		by increasing the rate to bind to the filament $u_{BF}$ from the baseline value \SI{100}{\micro\metre\per\second}
		to \SI{425}{\micro\metre\per\second} (yellow exponential line, 1),
		or set $v_{F}=0$ and vary $u_{BF}$ between \SI{100}{\micro\metre\per\second}
		and \SI{5250}{\micro\metre\per\second} (red vertical line, 2).
		All theoretical curves are compared to simulations with a bulk diffusion constant $D_{B}=\SI{60}{\micro\metre\squared\per\second}$ (points),
		confirming the agreement with the assumption $D_{B}\rightarrow \infty$.
		We see that the quality of the membrane polarization can be improved much more efficiently by dissipating power in the binding cycle
		than by transporting the proteins on the filament.
	}
\end{figure}
\begin{figure}
	\includegraphics{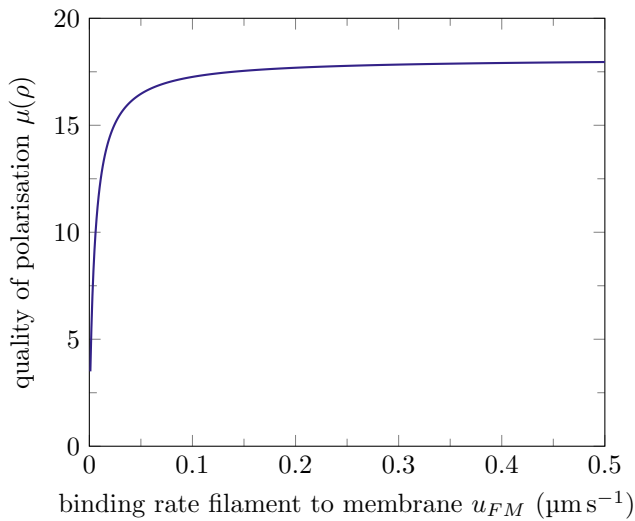}
	\caption{\label{fig:polarisation_vs_rate_filament_membrane}
		The quality of polarization as a function of the rate to move from the filament to the membrane, $u_{FM}$.
		All other parameter values are listed in Table~\ref{tab:polarisation_parameter_values}.
		Increasing $u_{FM}$ to a value larger than $v_{F}=\SI{0.5}{\micro\metre\per\second}$ has no influence on the high density spot on the membrane,
		only reducing the crowding at the end of the filament.
	}
\end{figure}
First, we vary the drift velocity on the filament $v_{F}$ between \num{0} and \SI{1}{\micro\metre\per\second},
and calculate both the dissipated power $P_{\mathrm{exact}}$ (see Eq.\,\ref{eq:dissipated_power_exact})
and the quality of the membrane spot $\mu\!\left(\rho\right)$ for each value of $v_{F}$.
The results of the exact solutions are plotted in Fig.\,\ref{fig:polarisation_vs_dissipation} (dark blue line)
together with the values obtained in simulations with $D_{B}=\SI{60}{\micro\metre\squared\per\second}$ (dark blue dots).
We see that the membrane spot becomes more pronounced if we increase $v_{F}$,
but that the dissipation per particle also increases strongly.
Combined, the quality of the membrane spot appears to increase with the dissipated power as a power law
when it is only driven by active transport.

Instead of increasing the drift velocity, the system may also break detailed balance in
the binding cycle filament-membrane-bulk-filament, lowering the value of the non-dimensional parameter $\delta<1$.
In Sec.\,\ref{sec:parameter_values_membrane_polarisation}, we noted that the rate to bind from the filament to the membrane $u_{FM}$
may be different from the rate to bind from the bulk to the membrane $u_{BM}$.
For example, the motor proteins that transport the particles on the filament may also provide a mechanism to deliver them to the membrane.
Furthermore, the chemical state of the proteins may be different when they are on the filament,
for example by strongly favoring the binding of (de)phosphorylated particles to the motor proteins,
or by packing the proteins on a vesicle before transport.
Hence, we vary the rate $u_{FM}$ while maintaining the driving velocity $v_{F}=\SI{0.5}{\micro\metre\per\second}$,
and calculate both $\mu\!\left(\rho\right)$ and $P_{\mathrm{exact}}$.
Fig.\,\ref{fig:polarisation_vs_dissipation} (light blue line and dots) shows that increasing $u_{FM}$ improves the quality of the polarized spot,
but it does not cost more free energy.
In fact, the power per particle caused by the binding cycle $P_{\mathrm{exact},FMB}$ becomes non-zero and remains small,
but increasing $u_{FM}$ also reduces the time that particles spend on the filament,
which reduces the probability that a particle unbinds from the filament to the bulk
and wastes the free energy that was spent in transporting it.
Hence, the power per particle caused by transport on the filament $P_{\mathrm{exact},F}$
decreases more strongly than $P_{\mathrm{exact},F}$ increases when we increase $u_{FM}$, leading to a lower overall power consumption.

The boundary condition Eq.\,\ref{eq:bc_f_0} indicates that the derivative of $f\!\left(z\right)$ at $z=0$,
the density on the filament near the membrane, becomes less negative or even positive when $u_{FM}$ increases.
Faster transmission of the particles from the filament to the membrane lowers the high density peak on the filament,
which is the spike at $z=0$ in Fig.\,\ref{fig:polarisation_filament_histogram}.
However, if $u_{FM}$ becomes much larger than $v_{F}$, the density $f\!\left(0\right)$ becomes zero,
and the flux to the membrane becomes limited by the speed at which new particles are transported by the motor proteins.
Fig.\,\ref{fig:polarisation_vs_rate_filament_membrane} shows that the quality of membrane polarization $\mu\!\left(\rho\right)$
reaches an asymptote before reaching a value of $u_{FM}\approx v_{F}$.
Even though increasing $u_{FM}$ increases polarization at almost no cost, it is only beneficial as long as $u_{FM}<v_{F}$.

To further investigate if increasing the dissipation in the binding cycle is efficient,
we choose to vary the rate to bind from the bulk to the filament, $u_{BF}$.
In Sec.\,\ref{sec:parameter_values_membrane_polarisation}, we showed that there is some freedom in choosing this rate.
We increase the rate from $u_{BF}=\SI{100}{\micro\metre\per\second}$ to \SI{350}{\micro\metre\per\second},
while setting $u_{FM}=v_{F}=\SI{0.5}{\micro\metre\per\second}$.
As shown in Fig.\,\ref{fig:polarisation_vs_dissipation} (yellow line, 1), the quality of polarization $\mu\!\left(\rho\right)$
increases exponentially with the power dissipated, which is more efficient than simply increasing the drift velocity on the filament.
The dissipation mainly increases because the number of particles on the filament increases with $u_{BF}$,
and this leads to a larger dissipation by the motor proteins.

To see if a non-equilibrium binding cycle can cause a polarized distribution of particles on the membrane,
we also vary $u_{BF}$ while keeping $v_{F}=0$.
Fig.\,\ref{fig:polarisation_vs_dissipation} (red line, 2) shows the model does show a strong polarization when only the binding cycle breaks detailed balance,
and that the free-energy dissipation is orders of magnitude lower than with a finite drift velocity.
For $u_{BF}=\SI{4300}{\micro\metre\per\second}$ and $v_{F}=0$,
we find $\mu\!\left(\rho\right) = 60$ and $P_{\mathrm{exact}} = \SI{0.02}{k_{B}T\per\second}$,
whereas for $u_{BF}=\SI{350}{\micro\metre\per\second}$ and $v_{F}=\SI{0.5}{\micro\metre\per\second}$,
we find the same polarization $\mu\!\left(\rho\right) = 60$ but with a dissipated power of $P_{\mathrm{exact}} = \SI{7.8}{k_{B}T\per\second}$.
Hence, by driving the binding cycle out of equilibrium, via e.g. a phosphorylation-dephosphorylation cycle as in the Pom1/Tea1/Tea4 system,
the cell can make the binding from the filament to the membrane more likely
without increasing the backward rate from the membrane to the filament.
This process is probably highly efficient in terms of its free-energy consumption.

\section{\label{sec:polarisation_discussion}Discussion}
Cytoskeletal filaments are often organized in non-homogeneous structures within the cell.
For example, the mitotic spindle is a structure in which microtubules point radially outward
from two microtubule organizing centers.
Together with the intrinsically polarized structure of the filaments,
the cell can use its cytoskeleton to provide directional transport of other particles and break the symmetry of the distribution of particles.
Fission yeast makes use of microtubule based transport of Tea1 and Tea4
to polarize the distribution of Pom1 on its membrane~\cite{hachet_phosphorylation_2011},
and budding yeast transports vesicles with membrane-bound Cdc42 along actin cables,
leading to a polarized distribution of Cdc42 on the outer membrane~\cite{harris_cdc42_2010,martin_cell_2014}.
Hence, if the filaments point towards the membrane and the transported cargoes bind to the membrane,
transport by motor proteins along filaments offers a mechanism to polarize a distribution of proteins on the membrane of the cell.
Using a minimal model, we showed that transport along a filament can create a polarized steady state distribution on the membrane,
and that the motor proteins dissipate a reasonable amount of chemical free energy to maintain this non-equilibrium state.
Using biologically relevant parameter values, the protein concentration on the membrane can increase by a factor \num{12}
compared to the equilibrium density on the membrane, forming a high density spot close to the filament
while the cell dissipates on average $\SI{2.6}{k_{B}T\per\second}$ of chemical free energy per particle.
In comparison, the hydrolysis of a single ATP molecule releases roughly $\SI{18}{k_{B}T}$~\cite{tran_changes_1998},
so a motor protein moving at \SI{0.5}{\micro\metre\per\second} that hydrolyses one ATP molecule per \SI{8}{\nano\metre} step
dissipates more than $\SI{1000}{k_{B}T\per\second}$.
Hence, a polarized distribution of filament orientations that is used for directed transport can be sufficient
to create a polarized distribution of cargo proteins on the membrane.

The model also includes a second mechanism to break detailed balance besides active transport.
The particles transition reversibly between the bulk (cytosol) and filament, between the filament and membrane,
and between the membrane and bulk.
Those reactions can break detailed balance by modifying the particles as they pass through the loop.
For example, fission yeast Pom1 is dephosphorylated before it binds to the membrane,
but its autophosphorylation then increases its dissociation rate from the membrane~\cite{hachet_phosphorylation_2011, martin_cell_2014}.
Furthermore, budding yeast Cdc42 switches between GDP and GTP bound states, binding to the membrane as Cdc42-GDP
and unbinding after Cdc24 and Bem1 exchanges the bound nucleotide for GTP~\cite{harris_cdc42_2010}.
The model shows that such a non-equilibrium binding cycle would dissipate around $\SI{7e-3}{k_{B}T\per\second}$ per particle
to generate the same level of polarization ($\mu\!\left(\rho\right)=12$) as active transport,
which dissipates $\SI{2.6}{k_{B}T\per\second}$ per particle.
Even though a driven binding cycle can thus dissipate orders of magnitude less free-energy
than active transport along the filament to create the same membrane spot,
the absolute dissipation of active transport could still be low compared to other cellular processes such as protein translation,
and therefore evade evolutionary selection pressure.
Previously, it was estimated that budding yeast experiences selection pressure against genes
that require more than \num{1e4} molecules of ATP per cell cycle~\cite{lynch_bioenergetic_2015}.
We can make an estimate of the energy costs of cell polarization via transport alone
by assuming that the polarization machinery is active for \SI{30}{\minute} per cell cycle,
which is a quarter of the minimal duration of the cell cycle of yeast~\cite{lynch_bioenergetic_2015}.
Furthermore, we assume that \num{1000} copies of the polarizing protein are involved,
similar to the copy number of Cdc42 in budding yeast~\cite{lahtvee_absolute_2017},
and that ATP hydrolysis releases $\SI{18}{k_{B}T}$~\cite{tran_changes_1998}.
Then, the power predicted by our model in the biologically relevant regime ($\SI{2.6}{k_{B}T\per\second}$) leads to the estimate that
active transport would hydrolyze around \num{2.6e5} molecules of ATP per cell cycle to create cell polarization.
Hence, ignoring other fitness effects of the different polarization mechanisms,
the energetic requirements alone can be sufficient to stimulate the formation of a driven binding cycle.

In fact, our work provides a new perspective on the design logic of these systems.
In particular, experiments indicate that Pom1 itself is not actively transported
along the microtubule filaments~\cite{hachet_phosphorylation_2011,martin_cell_2014}.
Instead, the filaments only mark the location where Pom1 is delivered to the membrane
by transporting Tea4 to the microtubule tip, specifying the position where the phosphatase Dis2 becomes active in dephosphosphorylating Pom1,
allowing Pom1 to bind to the membrane~\cite{hachet_phosphorylation_2011,martin_cell_2014}.
The filaments could in principle also be used to actively transport Pom1, but our work suggests that this would only marginally enhance polarization
while it would significantly increase energy dissipation.
When active transport of the polarizing protein does occur, this transport may be sufficient to create a polarized protein density on the membrane,
but it is still more efficient to implement an additional driven binding cycle,
which is the case for the Cdc42 system in budding yeast.

A non-equilibrium binding cycle is more efficient because the particles only dissipate free-energy when they participate in the flux
that moves from the filament to the membrane, and it is this flux that causes the spot on the membrane.
In contrast, the motor proteins drive proteins along the entire filament,
and many unbind from the filament before they reach the membrane.
These cycles waste the free energy that was invested by the motor proteins,
because the high diffusion constant in the bulk almost immediately homogenizes the protein distribution in the bulk again.
However, the binding cycle will likely also lead to the wasteful release of chemical free energy
that is not captured by our model.
For example, a phosphorylation cycle will typically include erroneous dephosphorylation steps,
dissipating the free energy obtained from hydrolyzing an NTP molecule.
Integrating such wasteful processes in the minimal model presented here,
it is likely that breaking detailed balance in the cycle of binding reactions also leads to a free-energy cost.
In addition, simulations show that the transport along filaments is less wasteful
when the finite magnitude of the diffusion constant in the bulk is taken into account,
bringing the efficiencies of transport and binding closer together.
Nonetheless, it is fundamentally difficult to create a membrane spot when binding is in equilibrium,
because detailed balance causes the binding rate from the bulk to the membrane to be large
when the binding rate from the filament to the membrane is large.
This increased binding rate from the bulk leads to a large homogeneous equilibrium concentration of proteins on the membrane,
lowering the relative effect of the proteins deposited by the filament.
Furthermore, we saw in Fig.\,\ref{fig:polarisation_filament_histogram} that transport generates a high density spot on the membrane
via a direct reversible interchange of particles between this membrane spot and a strongly compressed density of particles on the tip of the filament.
When exclusion effects between the proteins on the filament are taken into account,
it may not be possible to create such a high density concentration on the tip of the filament~\cite{leduc_molecular_2012,rank_crowding_2018},
preventing the formation of a membrane spot.
In Sec.\,\ref{sec:polarisation_free_energy_costs}, we show that increasing the rate to bind from the filament to the membrane
reduces the protein density at the front of the filament, but then an active binding cycle is required to still create a significant spot on the membrane.
Hence, it is theoretically possible to create a polarized distribution of proteins on the membrane by active transport alone,
but both in cells or in synthetic systems,
membrane polarization would likely be superior using a dissipative mechanism that biases the binding from the filament to the membrane.

\begin{acknowledgments}
	This work was supported by European Research Council (ERC) Synergy Grant 609822,
	is part of the research program of the Netherlands Organization for
	Scientific Research (NWO), and performed at the research institute AMOLF.
\end{acknowledgments}

\input{main.bbl}


\widetext
\pagebreak
\begin{center}
	\textbf{\large Supplemental Material: \\ 
		Energetic constraints on filament mediated cell polarization}
\end{center}

\setcounter{equation}{0}
\setcounter{figure}{0}
\setcounter{table}{0}
\setcounter{page}{1}
\setcounter{section}{0}
\makeatletter
\renewcommand{\theequation}{S.\arabic{equation}}
\renewcommand{\thesection}{S.\Roman{section}}
\renewcommand{\thefigure}{S.\arabic{figure}}
\renewcommand{\thetable}{S.\arabic{table}}
\renewcommand{\bibnumfmt}[1]{[S#1]}
\renewcommand{\citenumfont}[1]{S#1]}

%
%
\begin{table}[b]
	\begin{tabularx}{\textwidth}{XX}
		\hline\hline
		Filament radius $r$ & \SI{0.0125}{\micro\metre}\rule{0pt}{2.5ex}\\
		Container radius $R$ & \SI{10.0125}{\micro\metre}\\
		Filament length $L$ & \SI{10}{\micro\metre}\\
		Diffusion constant on filament $D_{F}$ & \SI{0.004}{\micro\meter\squared\per\second}\\
		Driving velocity on filament $v_{F}$ & \SI{0.5}{\micro\metre\per\second}\\
		Diffusion constant on membrane $D_{M}$ & \SI{5}{\micro\metre\squared\per\second}\\
		Diffusion constant in bulk $D_{B}$ & \SI{60}{\micro\metre\squared\per\second}\\
		Binding rate to the filament $u_{BF}$ & \SI{100}{\micro\metre\per\second}\\
		Unbinding rate from filament $k_{FB}$ & \SI{0.5}{\per\second}\\
		Binding rate to membrane $u_{BM}$ & \SI{0.01}{\micro\metre\per\second}\\
		Unbinding rate from membrane $k_{MB}$ & \SI{0.1}{\per\second}\\
		Rate of deposit on the membrane $u_{FM}$ & \SI{0.01}{\micro\metre\per\second}\\
		Reverse rate from the membrane $u_{MF}$ & \SI{20}{\micro\metre\per\second}\\
		\hline\hline
	\end{tabularx}
	\caption
	{\label{tab:polarisation_parameter_values}
		Model parameters and values, as derived in Sec.\,\ref{sec:parameter_values_membrane_polarisation}.
		For most parameters, we found a biologically relevant value in the literature,
		while for $u_{BF}$ we chose a value that allows for the formation of a pronounced membrane spot,
		and $u_{MF}$ is set by detailed balance.
	}
\end{table}
\begin{table}[b]
	\begin{tabularx}{\textwidth}{XX}
		\hline\hline
		Filament radius $\rho$ & \num{1.8e-3}\rule{0pt}{2.5ex}\\
		Container radius $\Rho$ & \num{1.4}\\
		Filament length $\Lambda$ & \num{112}\\
		Drift velocity $\alpha$ & \num{11}\\
		Filament-membrane transition rate $\beta$ & \num{0.22}\\
		Membrane-filament transition rate $\gamma$ & \num{28}\\
		Binding driven factor $\delta$ & \num{1}\\
		\hline\hline
	\end{tabularx}
	\caption
	{\label{tab:polarisation_dimensionless_parameter_values}
		The values of the dimensionless parameters, defined in Eq.\,\ref{eq:nondimensionalisation_parameters},
		given by the values listed in Table~\ref{tab:polarisation_parameter_values}.
	}
\end{table}
\begin{table}[b]
	\begin{tabularx}{\textwidth}{XX}
		\hline\hline
		Longitudinal reaction length $l_{\mathrm{long}}$ & \SI{5e-3}{\micro\metre}\rule{0pt}{2.5ex}\\
		Radial reaction length $l_{\mathrm{rad}}$ & \SI{5e-3}{\micro\metre}\\
		Time step $\delta t$ & \SI{1e-6}{\second}\\
		\hline\hline
	\end{tabularx}
	\caption
	{\label{tab:polarisation_simulation_parameter_values}
		The parameter values used for the simulations of a system with a finite bulk diffusion constant.
		The reaction lengths were chosen such that they are (much) smaller than all length scales in the system,
		but large enough that a diffusing particle with diffusion constant $D_{B}$ that comes close to the membrane
		has a significant probability to be within a reaction volume during a time step $\delta t$.
		The time step was chosen such that all reaction rates lead to a small (<1\%) probability of performing a reaction each time step.
	}
\end{table}
\section{\label{sec:polarisation_solution_integration_constants}Integration constants of analytical solutions}
In the Eq.\,\ref{eq:ss_phi_dif} and Eq.\,\ref{eq:ss_mu_dif}, we found two ordinary differential equations for the non-dimensionalized
protein densities on the filament and on the membrane.
These equations are valid in steady state, and when the diffusion constant in the bulk $D_{B}$ is very large,
such that there are no density fluctuations in the bulk.
The solutions for the dimensionless density on the filament $\varphi\!\left(\lambda\right)$
and the dimensionless density on the membrane $\mu\!\left(\sigma\right)$
are given in Eq.\,\ref{eq:phi_general_sol} and Eq.\,\ref{eq:mu_general_sol}, respectively.
These solutions involve the integration constants $C_{1}$, $C_{2}$, $C_{3}$, and $C_{4}$,
which are determined by the four boundary conditions Eqs.\,\ref{eq:ss_bc_phi_0}--\ref{eq:ss_bc_mu_Rho}.
The boundary conditions provide a set of linear equations on the integration constants,
which can be solved by standard linear algebra.
Using that the modified Bessel functions obey the relations
\begin{align}
	\partial_{\sigma} I_{0} \left(\sigma \right) &= I_{1} \left( \sigma \right), \label{eq:derivative_bessel_I} \\
	\partial_{\sigma} K_{0} \left(\sigma\right) &= -K_{1} \left( \sigma \right), \label{eq:derivative_bessel_K}
\end{align}
we find the following integration constants,
\begin{align}
	C_{1} &= \frac{1}{\mathcal{N}} \Bigg\{
	\left( I_{1}\left(\Rho\right) K_{1} \left( \rho \right) - K_{1} \left( \Rho \right) I_{1}\left(\rho\right) \right) \nonumber \\
	&\cdot \bigg[ \alpha \left(\sqrt{4+\alpha^{2}}+\alpha \right)
	\left(1- \exp \left[ \frac{\Lambda}{2} \left(\sqrt{4+\alpha^{2}}-\alpha \right) \right] \right) \nonumber \\
	&\quad+ \beta \left( 1-\delta \right) \left(\sqrt{4+\alpha^{2}}+\alpha \right)
	\exp \left[ \frac{\Lambda}{2} \left(\sqrt{4+\alpha^{2}}-\alpha \right) \right] -2 \alpha \beta \bigg] \nonumber \\
	& + \alpha \gamma \left( I_{1}\left(\Rho\right) K_{0} \left( \rho \right)
	+ K_{1} \left( \Rho \right) I_{0}\left(\rho\right) \right) \nonumber \\
	&\cdot \left(\sqrt{4+\alpha^{2}}+\alpha \right)
	\left(1- \exp \left[ \frac{\Lambda}{2} \left(\sqrt{4+\alpha^{2}}-\alpha \right) \right] \right) \Bigg\}, \label{eq:C1}
\end{align}
\begin{align}
	C_{2} &= \frac{1}{\mathcal{N}} \Bigg\{
	\left( I_{1}\left(\Rho\right) K_{1} \left( \rho \right) - K_{1} \left( \Rho \right) I_{1}\left(\rho\right) \right) \nonumber \\
	&\cdot \bigg[ \alpha \left(\sqrt{4+\alpha^{2}}-\alpha \right)
	\left(1- \exp \left[ -\frac{\Lambda}{2} \left(\sqrt{4+\alpha^{2}}+\alpha \right) \right] \right) \nonumber \\
	&\quad+ \beta \left( 1-\delta \right) \left(\sqrt{4+\alpha^{2}}-\alpha \right)
	\exp \left[ -\frac{\Lambda}{2} \left(\sqrt{4+\alpha^{2}}+\alpha \right) \right] +2 \alpha \beta \bigg] \nonumber \\
	& + \alpha \gamma \left( I_{1}\left(\Rho\right) K_{0} \left( \rho \right)
	+ K_{1} \left( \Rho \right) I_{0}\left(\rho\right) \right) \nonumber \\
	&\cdot \left(\sqrt{4+\alpha^{2}}-\alpha \right)
	\left(1- \exp \left[ -\frac{\Lambda}{2} \left(\sqrt{4+\alpha^{2}}+\alpha \right) \right] \right) \Bigg\}, \label{eq:C2}
\end{align}
\begin{align}
	C_{3} &= \frac{\gamma \, I_{1} \left(\Rho \right)}{\delta \, \mathcal{N}} \Bigg\{ \left( 1-\delta \right) \nonumber \\
	&\cdot \left(-2 \exp \left[ \frac{\Lambda}{2} \left(\sqrt{4+\alpha^{2}}-\alpha \right) \right]
	+ 2 \exp \left[ -\frac{\Lambda}{2} \left(\sqrt{4+\alpha^{2}}+\alpha \right) \right] \right) \nonumber \\
	& -\alpha \Big[\left(\sqrt{4+\alpha^{2}}+\alpha \right)
	\exp \left[ \frac{\Lambda}{2} \left(\sqrt{4+\alpha^{2}}-\alpha \right) \right] \nonumber \\
	&\qquad + \left(\sqrt{4+\alpha^{2}}-\alpha \right)
	\exp \left[ -\frac{\Lambda}{2} \left(\sqrt{4+\alpha^{2}}+\alpha \right) \right] \Big]
	+2 \alpha \sqrt{4+\alpha^{2}} \Bigg\}, \label{eq:C3}
\end{align}
\begin{align}
	C_{4} &= \frac{\gamma \, K_{1} \left(\Rho \right)}{\delta \, \mathcal{N}} \Bigg\{ \left( 1-\delta \right) \nonumber \\
	&\cdot \left(-2 \exp \left[ \frac{\Lambda}{2} \left(\sqrt{4+\alpha^{2}}-\alpha \right) \right]
	+ 2 \exp \left[ -\frac{\Lambda}{2} \left(\sqrt{4+\alpha^{2}}+\alpha \right) \right] \right) \nonumber \\
	& -\alpha \Big[\left(\sqrt{4+\alpha^{2}}+\alpha \right)
	\exp \left[ \frac{\Lambda}{2} \left(\sqrt{4+\alpha^{2}}-\alpha \right) \right] \nonumber \\
	&\qquad + \left(\sqrt{4+\alpha^{2}}-\alpha \right)
	\exp \left[ -\frac{\Lambda}{2} \left(\sqrt{4+\alpha^{2}}+\alpha \right) \right] \Big]
	+2 \alpha \sqrt{4+\alpha^{2}} \Bigg\}. \label{eq:C4}
\end{align}
To prevent repetition, we have defined the denominator $\mathcal{N}$,
\begin{align}
	\mathcal{N} &= \left( I_{1}\left(\Rho\right) K_{1} \left( \rho \right) - K_{1} \left( \Rho \right) I_{1}\left(\rho\right) \right)
	\Bigg\{ \nonumber \\
	&\quad-2 \exp \left[ \frac{\Lambda}{2} \left(\sqrt{4+\alpha^{2}}-\alpha \right) \right]
	+ 2 \exp \left[ -\frac{\Lambda}{2} \left(\sqrt{4+\alpha^{2}}+\alpha \right) \right] \nonumber \\
	&\quad - \beta \bigg[ \left(\sqrt{4+\alpha^{2}}+\alpha \right)
	\exp \left[ \frac{\Lambda}{2} \left(\sqrt{4+\alpha^{2}}-\alpha \right) \right] \nonumber \\
	&\qquad \quad + \left(\sqrt{4+\alpha^{2}}-\alpha \right)
	\exp \left[ -\frac{\Lambda}{2} \left(\sqrt{4+\alpha^{2}}+\alpha \right) \right] \bigg] \Bigg\} \nonumber \\
	&+ \gamma \left( I_{1}\left(\Rho\right) K_{0} \left( \rho \right) + K_{1} \left( \Rho \right) I_{0}\left(\rho\right) \right)
	\Bigg\{ \nonumber \\
	&\quad-2 \exp \left[ \frac{\Lambda}{2} \left(\sqrt{4+\alpha^{2}}-\alpha \right) \right]
	+ 2 \exp \left[ -\frac{\Lambda}{2} \left(\sqrt{4+\alpha^{2}}+\alpha \right) \right] \Bigg\}. \label{eq:norm}
\end{align}
Using these definitions and Eq.\,\ref{eq:phi_general_sol} and Eq.\,\ref{eq:mu_general_sol},
we now have a full analytical solution for the protein densities on the filament and the membrane.

\section{\label{sec:reflection_algorithms}Reflecting boundary algorithms}
The Monte Carlo algorithm that simulates the diffusion of the particle in each of the system parts
makes a finite spatial step every time step.
If the particle is at the longitudinal position $z_{0}$ in the bulk or on the filament,
then the algorithm proposes a new position $z_{1}'$ by taking a step $\delta z$,
\begin{equation}
	z_{1}' = z_{0} + \delta z,
\end{equation}
where the difference is drawn from a normal distribution, $\delta z \sim \mathcal{N}\!\left(0,\sqrt{2 D \delta t}\right)$
and $D$ is the appropriate diffusion constant.
If this proposed position $z_{1}'$ is outside the container, it is reflected back in the boundary.
Assuming that $L \gg \sqrt{2 D \delta t}$, such that a particle is never reflected twice, we can summarize the diffusive algorithm by
\begin{align}
	z_{1} &= z_{1}' &\mathrm{if} \ 0 \leq z_{1}' \leq L, \\
	z_{1} &= -z_{1}' &\mathrm{if} \ z_{1}'<0, \\
	z_{1} &= 2L - z_{1}' &\mathrm{if} \ z_{1}' > L.
	\label{eq:reflect_longitudinal}
\end{align}
These reflections are reversible.
For example, a particle that is close to $z=0$ can move from $z_{0}>0$ to $z_{1}>0$ directly,
or it can move to $z_{1}'=-z_{1}$ after which it is reflected to $z_{1}$.
If we denote the probability density function of $\mathcal{N}\!\left(0,\sqrt{2 D \delta t}\right)$ by $n\!\left(z\right)$,
then the total probability density for the transition equals
\begin{equation}
	p\!\left(z_{0} \rightarrow z_{1} \right)= n\!\left(z_{1}-z_{0}\right) + n\!\left(z_{1}' - z_{0}\right)
	= n\!\left(z_{0}-z_{1}\right) + n\!\left(z_{0}' - z_{1}\right) = p\!\left(z_{1} \rightarrow z_{0}\right),
	\label{eq:prob_density_step_0_to_1}
\end{equation}
where we used that $z_{1}'=-z_{1}$ and $z_{0}'=-z_{0}$.
Since the probability density function of the normal distribution only depends on the traversed distance
and not on the direction (sign) of the step, and because reflections preserve distance, the reflective algorithm obeys detailed balance.

\begin{figure}
	\includegraphics{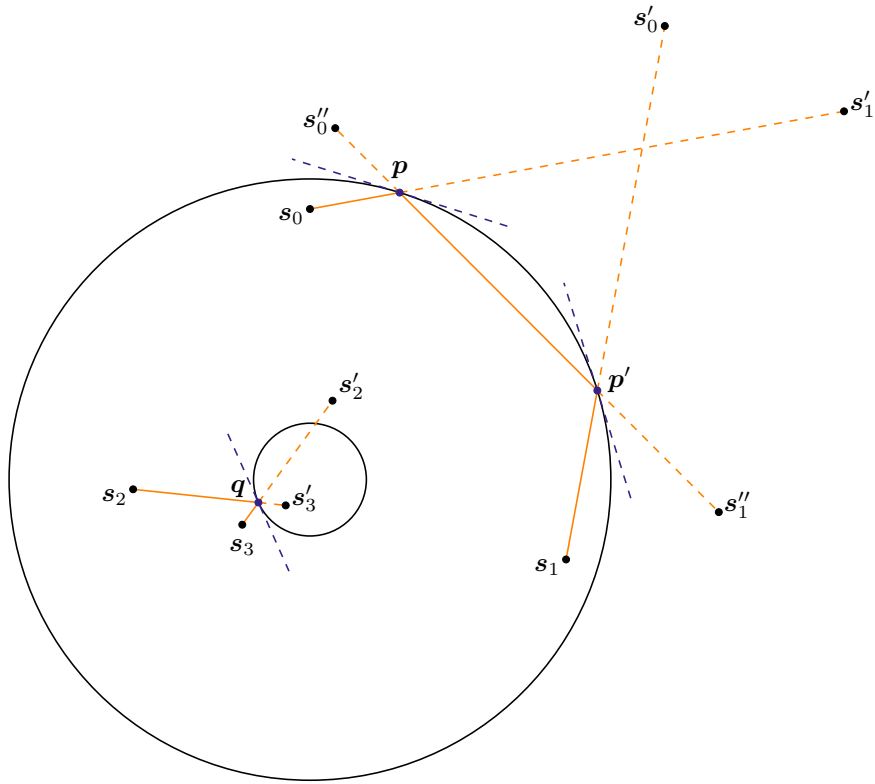}
	\caption{\label{fig:2d_reflection}
		Starting from a position $\boldsymbol{s}_{0}$, the diffusion algorithm can propose a new point $\boldsymbol{s}_{1}'$ outside of the container.
		To bring the proposed point back into the outer circle, we find the point $\boldsymbol{p}$
		where the line $\boldsymbol{s}_{0}\text{---}\boldsymbol{s}_{1}'$ intersects the outer circle,
		and reflect $\boldsymbol{s}_{1}'$ in the line tangent to the circle through $\boldsymbol{p}$.
		The resulting point, $\boldsymbol{s}_{1}''$, still lies outside the circle in this example,
		so we repeat the procedure, reflecting $\boldsymbol{s}_{1}''$ in the tangent line through $\boldsymbol{p}'$, giving $\boldsymbol{s}_{1}$.
		Starting from $\boldsymbol{s}_{1}$, taking the proposed point $\boldsymbol{s}_{0}'$ leads to the exact reverse of the first path.
		Similarly, we can start from a position $\boldsymbol{s}_{2}$ and propose a point $\boldsymbol{s}_{3}'$ that lies inside the inner circle.
		Again, we find the intersection point $\boldsymbol{q}$ and reflect the point in the tangent line through $\boldsymbol{q}$ to find $\boldsymbol{s}_{3}$.
		The reverse transition from $\boldsymbol{s}_{3}$ to $\boldsymbol{s}_{2}$ occurs when we propose a point $\boldsymbol{s}_{2}'$
		that is outside of the inner circle, showing that we have to reflect any line that passes through the inner circle to guarantee reversibility of all paths.
	}
\end{figure}
When the particle is in the bulk or on the membrane, it makes two dimensional diffusive steps in the $x-y$ plane,
where the radial coordinate $s=\sqrt{x^{2} + y^{2}}$.
Starting from a position $\boldsymbol{s}_{0}$, a new position $\boldsymbol{s}_{1}'=\boldsymbol{s}_{0}+\delta\boldsymbol{s}$ is proposed,
where both Cartesian coordinates of the difference vector $\delta \boldsymbol{s}$
are drawn from a normal distribution with standard deviation $\sqrt{2 D \delta t}$.
The proposed position can be inside the inner circle that represents the filament, $\left \lVert \boldsymbol{s}_{1}' \right \rVert < r$,
or outside the outer circle that borders the container, $\left \lVert \boldsymbol{s}_{1}' \right \rVert > R$.
As shown in Fig.\,\ref{fig:2d_reflection}, a path that crosses the outer circle is reflected in tangent line to the circle.
To find the reflection point $\boldsymbol{p}$, we define points along the line $\boldsymbol{s}_{0}\text{---}\boldsymbol{s}_{1}'$ as
\begin{equation}
	\boldsymbol{l}\!\left(\lambda\right) =\boldsymbol{s}_{0} + \lambda \left(\boldsymbol{s}_{1}'-\boldsymbol{s}_{0}\right),
	\label{eq:definition_line_outer_reflection}
\end{equation}
where $\boldsymbol{s}_{0} = \boldsymbol{l}\!\left(0\right)$ and $\boldsymbol{s}_{1}'=\boldsymbol{l}\!\left(1\right)$.
Then, $\boldsymbol{p}$ is the point where this line intersects the outer circle, providing the quadratic equation
\begin{equation}
	\left\lVert\boldsymbol{l}\!\left(\lambda\right)\right\rVert^{2} = R^{2}.
	\label{eq:definition_line_outer_circle_crossing}
\end{equation}
Then, the reflection point is given by the positive root of this equation, $\boldsymbol{p} = \boldsymbol{l}\!\left(\lambda_{+}\right)$, with
\begin{equation}
	\lambda_{+} = \frac{1}{\left\lVert \boldsymbol{s}_{1}'-\boldsymbol{s}_{0}\right\rVert^{2}}
	\left[\sqrt{\left(\boldsymbol{s}_{0}\cdot\left(\boldsymbol{s}_{1}'-\boldsymbol{s}_{0}\right)\right)^{2}
		+\left(R^{2}-\left\lVert \boldsymbol{s}_{0}\right\rVert^{2}\right)\left\lVert \boldsymbol{s}_{1}'-\boldsymbol{s}_{0}\right\rVert^{2}}
	-\boldsymbol{s}_{0}\cdot\left(\boldsymbol{s}_{1}'-\boldsymbol{s}_{0}\right)\right].
	\label{eq:solution_line_outer_circle_crossing}
\end{equation}
Knowing $\boldsymbol{p}$, we define the unit vector $\widehat{\boldsymbol{p}}=\boldsymbol{p}/R$.
The proposed point $\boldsymbol{s}_{1}'$ is reflected by moving it in the direction of $\widehat{\boldsymbol{p}}$,
\begin{equation}
	\boldsymbol{r}_{1}'' = \boldsymbol{r}_{1}'  -2 \left[\widehat{\boldsymbol{p}}\cdot\boldsymbol{r}_{1}'-R\right] \widehat{\boldsymbol{p}}.
	\label{eq:reflected_position_outer}
\end{equation}
As shown in Fig.\,\ref{fig:2d_reflection}, it is not guaranteed that $\left\lVert\boldsymbol{r}_{1}''\right\rVert \leq R$,
so the algorithm checks whether the newly proposed point is inside the container, and if not it repeats the previous steps
replacing the old point $\boldsymbol{r}_{0}$ with $\boldsymbol{p}$ and replacing the proposed point $\boldsymbol{r}_{1}'$ with $\boldsymbol{r}_{1}''$.
The algorithm ends when the proposed point is inside the container, setting $\boldsymbol{r}_{1}$.

Fig.\,\ref{fig:2d_reflection} shows that we perform similar reflections in the inner circle with radius $r$ that represents the filament.
Starting from an initial position $\boldsymbol{s}_{2}$ inside the container, a proposed point $\boldsymbol{s}_{3}'$ can be inside the filament.
Furthermore, the line $\boldsymbol{s}_{2}\text{---}\boldsymbol{s}_{3}'$ can cross the inner circle
even when the proposed point is outside the filament.
In both cases, we reflect the line in the first point that crosses the inner circle.
Defining the reflection point $\boldsymbol{q}$ as the first intersection point between the inner circle and the line
\begin{equation}
	\boldsymbol{l}\!\left(\lambda\right) = \boldsymbol{s}_{2} + \lambda \left(\boldsymbol{s}_{3}'-\boldsymbol{s}_{2}\right),
	\label{eq:definition_line_inner_reflection}
\end{equation}
the reflection point follows from the smallest solution of the quadratic equation
\begin{equation}
	\left\lVert \boldsymbol{l}\!\left(\lambda\right)\right\rVert^{2} = r^{2}.
	\label{eq:definition_line_inner_circle_crossing}
\end{equation}
To test whether a reflection is necessary, we first check if
$\left\lVert \boldsymbol{s}_{3}'-\boldsymbol{s}_{2}\right\rVert > \left\lVert \boldsymbol{s}_{2}\right\rVert - r$.
If that is not the case, then the line $\boldsymbol{s}_{2}\text{---}\boldsymbol{s}_{3}'$ is not long enough to reach the circle.
If this test is positive, we calculate the discriminant of the second order polynomial in $\lambda$ given by Eq.\,\ref{eq:definition_line_inner_circle_crossing}.
If the discriminant is negative or zero, then the extended line through $\boldsymbol{s}_{2}$ and $\boldsymbol{s}_{3}'$
never crosses the inner circle or only touches it in one point, respectively.
In either case, no reflection is necessary.
If a reflection is still possible at that point, we calculate the smallest root of Eq.\,\ref{eq:definition_line_inner_circle_crossing},
\begin{equation}
	\lambda_{-} = \frac{1}{\left\lVert\boldsymbol{s}_{3}'-\boldsymbol{s}_{2}\right\rVert^{2}}
	\left[-\sqrt{\left(\boldsymbol{s}_{2} \cdot \left(\boldsymbol{s}_{3}'-\boldsymbol{s}_{2}\right)\right)^{2}
		-\left(\left\lVert \boldsymbol{s}_{2}\right\rVert^{2} - r^{2} \right) \left\lVert \boldsymbol{s}_{3}'-\boldsymbol{s}_{2} \right\rVert^{2}}
	-\boldsymbol{s}_{2} \cdot \left(\boldsymbol{s}_{3}'-\boldsymbol{s}_{2}\right)\right].
	\label{eq:solution_line_inner_circle_crossing}
\end{equation}
Finally, we test if $\lambda_{-}<0$ or if $\lambda_{-}\geq1$, which together with the assumption that $\left\lVert\boldsymbol{s}_{2}\right\rVert\geq r$
implies that the line $\boldsymbol{s}_{2}\text{---}\boldsymbol{s}_{3}'$ never crosses the circle.
If $0 \leq \lambda_{-} < 1$, we know that the line has crossed the circle, and we perform a reflection
in the point $\boldsymbol{q}=\boldsymbol{l}\!\left(\lambda_{-}\right)$.
Defining the unit vector $\widehat{\boldsymbol{q}} = \boldsymbol{q}/r$, the reflected point is located at
\begin{equation}
	\boldsymbol{s}_{3} = \boldsymbol{s}_{3}' + 2 \left[r - \widehat{\boldsymbol{q}}\cdot \boldsymbol{s}_{3}' \right] \widehat{\boldsymbol{q}}.
	\label{eq:reflected_position_inner}
\end{equation}
This reflection is guaranteed to lie outside of the inner circle, completing our diffusive algorithm.


\input{supplemental.bbl}

\end{document}

%% file: main.bbl
%

%% file: supplemental.bbl
%